\documentclass[12pt,preprint]{aastex}

\shorttitle{}
\shortauthors{}

\usepackage{graphicx}
\usepackage{amssymb}
\usepackage{amsmath}

\newcommand{\hi}{{\sc Hi} }

\newcommand{\kms}{km~s$^{-1}$ }
\newcommand{\kmsp}{km~s$^{-1}$}
\newcommand{\hii}{{\sc Hii} }
\newcommand{\AaA}{A\&A}
\newcommand{\ApJ}{ApJ}
\newcommand{\AJ}{AJ}
\newcommand{\JFM}{J. Fluid Mech.}
\newcommand{\MNRAS}{MNRAS}

\begin{document}

\title{On the use of fractional Brownian motion simulations to determine the 3D statistical properties of interstellar gas.}

\author{M.-A. Miville-Desch\^enes}
\affil{Canadian Institute for Theoretical Astrophysics, 60 St-George st, Toronto, Ontario, M5S 3H8, Canada}

\author{F. Levrier and E. Falgarone}
\affil{Laboratoire de Radioastronomie millim\'etrique, LERMA \& CNRS FRE 2460, \'Ecole Normale Sup\'erieure, 24 rue Lhomond, 75231, 
Paris Cedex 05, France}

\begin{abstract}
Based on fractional Brownian motion (fBm) simulations of 3D gas density and velocity fields, we present a study of the 
statistical properties of spectro-imagery observations (channel maps, integrated emission, and line centroid velocity) in the case of 
an optically thin medium at various temperatures. The power spectral index $\gamma_W$ of the integrated emission is identified with 
that of the 3D density field ($\gamma_n$) provided the medium's depth is at least of the order of the largest transverse scale in the image, 
and the power spectrum of the centroid velocity map is found to have the same index $\gamma_C$ as that of the 
velocity field ($\gamma_v$). Further tests with non-fBm density and velocity fields show that this last result holds, and is not modified either by the effects of 
density-velocity correlations. A comparison is made with the theoretical predictions of \citet{lazarian2000}. 
\end{abstract}


\section{Introduction}

The early works on interstellar turbulence, summarized by \citet{chandrasekhar49},  
pointed out that the interstellar medium (ISM) exhibits Reynolds numbers so large that it is very likely to be turbulent.
For  incompressible turbulent flows the \citet{kolmogorov41} theory predicts 
a dissipation-free energy cascade between large scales where turbulent energy is injected, down 
to small scales where it is dissipated and heats the gas.
In this inertial range the theory predicts a three-dimensional power law energy spectrum  
($E(k) \propto k^{-5/3}$), which has been observed extensively in terrestrial turbulence 
experiments (see \citet{grant62} for an example).

Since the work of \citet{chandrasekhar49}, several studies have shown 
that the interstellar medium density and velocity structures projected onto the plane of the sky 
are self-similar and well described by power laws. 
For instance, a number of works were done on H$\alpha$ centroid 
velocity fields in {\sc Hii} regions using auto-correlation and structure functions 
(see \citet{miville-deschenes95} and references therein). The self-similar structure
of  {\sc Hi} has also been studied, almost exclusively using the power spectrum of the 21 cm 
emission in galactic \citep{baker73,crovisier83,green93,deshpande2000,dickey2001} 
and extra-galactic regions \citep{spicker88,elmegreen2001,stanimirovic2001}.
Various studies aimed at describing the fractal properties of dust infrared emission 
were also done using power spectra \citep{gautier92,wright98}
and the area-perimeter relation \citep{bazell88,dickman90,vogelaar94}.
But it is probably on molecular clouds (CO emission) that most of the work has been done
using several statistical tools: the size-linewidth relation \cite[]{larson81,myers83,falgarone90},
the auto-correlation function  \citep[]{kleiner84,kleiner85,kleiner87,perault86,kitamura93,miesch94,larosa99},
wavelet decomposition \citep{gill90}, 
the area-perimeter relation \citep{falgarone91}, 
principal component analysis \citep{heyer97,brunt2002}
and the $\Delta$-variance method \citep{stutzki98,bensch2001}.

All those studies suggest that turbulence, for which such self-similar properties are expected, may play a key role 
in the structuration of the interstellar medium, both in density and velocity. 
Furthermore, by providing non-thermal support against 
self-gravitational collapse, turbulence also regulates star formation \citep[]{klessen2000,heitsch2001}. 
It might also have a strong and very non-linear impact on dust evolution by accelerating
the coagulation or fragmentation of grains \citep[]{falgarone95,miville-deschenes2002}, and on the ISM chemical evolution as intermittent energy dissipation could locally trigger 
endothermic reactions \citep[]{joulain98}. However, following \citet{scalo87}, interstellar turbulence 
may be very different from Kolmogorov's prediction. 
The interstellar medium is compressible and magnetized,
and several processes inject energy at very different scales 
(from galactic differential rotation down to stellar winds). 
Therefore, the description of interstellar turbulence in various environments is mandatory
to understand its role on the structure, kinematics, energy transfer, thermodynamics and chemistry of the gas at various scales.

One important issue here is to relate the statistical properties of the observed
quantities (integrated emission, channel maps and centroid velocity) to the
three-dimensional properties of the density and velocity fields, which actually describe
interstellar turbulence. This calls for reliable methods to extract the relevant information 
from spectro-imagery observations, called position-position-velocity (PPV) data cubes, 
and disentangle the density, velocity, and temperature contributions to the observed space-velocity structures.
There is also a need to understand the effects of 3D-2D projection and of line opacity 
on the observed statistical properties. 
Lately there has been a number of works based on PPV data cubes to deduce the statistical 
properties of the underlying three-dimensional density and velocity fields of \hi observations 
\citep[]{stanimirovic2001,elmegreen2001,dickey2001}. Most of these works are based on 
the theoretical work of \citet{lazarian2000} (hereafter LP00), who proposed a method to deduce the 3D density and 
velocity power spectral indexes, respectively noted $\gamma_n$ and $\gamma_v$ thereafter, 
from the PPV cubes.

In this paper, we study the spectral properties of simulated PPV cubes using 
fractional Brownian motion (fBm) density and velocity fields, with the twofold purpose of comparing 
them with the predictions of LP00, and proposing another method, based on the centroid velocity map, to 
determine the spectral index of the three-dimensional velocity field. The reason for using fBms, 
despite their lack of physical reality, lies with their well-behaved 
statistical properties and the ease with which they can be generated. With single-index power 
law power spectra and Gaussian fluctuations, fractional 
Brownian motions represent useful and flexible test cases for structure statistics 
retrieval methods. Such random fields were used for instance by \citet{stutzki98} to describe 
molecular cloud images using $\Delta$-variance analysis, and by \citet{brunt2002a} to retrieve 
the statistics of turbulent interstellar velocity fields through the use of Principal Component Analysis (PCA). 
It should however be stressed that these fields do not arise from physical simulations. 
In particular, they do not comply with hydrodynamics nor gravity, and are used solely for their statistical simplicity.

This kind of study based on fBms usefully complements those conducted on magnetohydrodynamical (MHD) simulations (see for instance 
\citet{ostriker2001}) because the lack of physics is somewhat compensated by the broad range of scales accessible to the analysis in terms 
of power laws. In numerical simulations of turbulence the inertial range is severely limited by the available resolution and the large number of 
scales affected by injection and dissipation.

The paper is organised as follows: In section~\ref{sec_fbm}, 
we briefly present fBms and their properties, in the context of designing simple model distributions for 
our three-dimensional density and velocity fields. The computation of PPV data cubes from these fields is 
detailed in section~\ref{sec_simul_ppv}, then an analysis of 
their spectral properties is done in section~\ref{sec_stat_prop}. Section~\ref{sec_conclu} presents our concluding remarks. 
An appendix contains some details on analytical computations.

\section{Density and velocity fields}

\label{sec_fbm}

\subsection{Definition and construction of fBms}

A fractional Brownian motion (fBm) is a Gaussian random field $F$ which can be defined, as done in \citet{voss85}, in any Euclidean dimension $N$ by the relation
\begin{equation}
\label{eq_def_fbm}
<[F({\bf r_2})-F({\bf r_1})]^2> ~\propto~||{\bf r_2}-{\bf r_1}||^{2H}
\end{equation}
where $H$, called the Hurst exponent, is a real number in $[0,1]$, and the brackets stand for a spatial average over positions ${\bf r_1}$ and ${\bf r_2}$. The value $H=0.5$ gives 
the usual Brownian motion, for which it is well known that the mean squared increment $<\Delta F^2>=<[F({\bf r_2})-F({\bf r_1})]^2>$ scales as the separation $||\Delta {\bf r}||=||{\bf r_2}-{\bf r_1}||$. This relation states that the fluctuations of the field are isotropic, and that their amplitude as a function of scale is a power law. It easily translates into a similar relation for the power spectrum, which is the Fourier transform 
of the autocorrelation function ${\cal{A}}_F({\boldsymbol{\rho}})=<F({\bf r})F({\bf r}+{\bf \boldsymbol{\rho}})>$:
\begin{equation}
\label{eq_fourier_fbm}
P_F({\bf k})=\tilde{{\cal{A}}}_F({\bf k}) \propto k^{\gamma}
\end{equation}
where isotropy is expressed by the fact that $P_F$ depends only on the wavenumber $k$, which is the length of the $N$-dimensional wave vector ${\bf k}$. 
The power spectral index $\gamma$ is related to the Hurst exponent and the dimension by $\gamma=-2H-N$. 

Several methods are available to numerically generate fBms 
(for instance as a collection of clumps with a power law mass spectrum, see \citet{stutzki98}), but the easiest is based on Eq.~\ref{eq_fourier_fbm} 
and the fact that $P_F$ is proportional to $|\tilde{F}|^2$, the squared amplitude of the Fourier transform of $F$. 
Hence we first compute an isotropic amplitude following a power law:
\begin{equation}
A({\bf k})=A_0k^{\gamma/2}
\end{equation}
where $A_0$ is a normalization factor. Then the phase $\phi({\bf k})$ 
is constructed using a random number generator algorithm based on a uniform distribution between $-\pi$ and $\pi$, and to make sure the image in direct space 
is real, we enforce the constraint $\phi(-{\bf k})=-\phi({\bf k})$. From the amplitude and the phase, the real and imaginary parts of the Fourier transform are 
computed:
\begin{equation}
\mathrm{Re}[\tilde{F}({\bf k})]=A({\bf k})\cos[\phi({\bf k})]
\end{equation}
\begin{equation}
\mathrm{Im}[\tilde{F}({\bf k})]=A({\bf k})\sin[\phi({\bf k})]
\end{equation}
and the fBm $F({\bf r})$ is simply obtained by taking the inverse Fourier transform of $\tilde{F}({\bf k})$.

\begin{figure}
\plotone{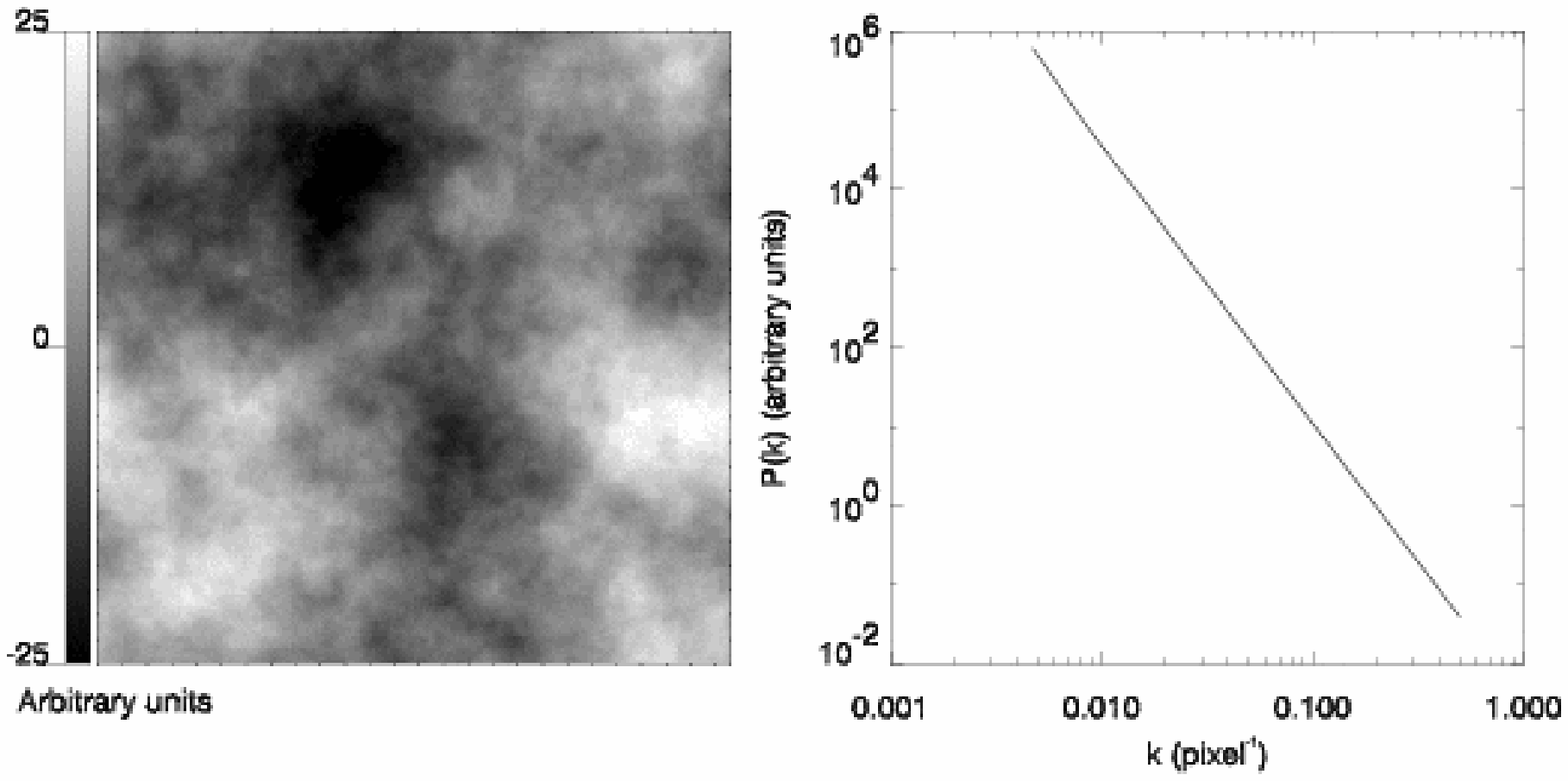}
\caption{\label{fig_example_fbm} Typical two-dimensional 257 $\times$ 257 fractional Brownian motion image with its power spectrum, which by 
construction follows a power law with an index equal to -3.5.}
\end{figure}

The method described here is general for any dimension, and the power spectra of the fields generated this way obey exact 
power laws\footnote{The only random elements are the phases.}, 
as can be seen in Fig.~\ref{fig_example_fbm}, which shows a two-dimensional (2D) $257 \times 257$ fBm image, 
along with its power spectrum\footnote{The odd size is simply a matter of convenience when building the fields. The Fourier transform algorithm is a bit slower, but speed was not critical in our simulations.}. 
In this figure we used a scatter plot (all points of the 2D amplitude are shown)
in order to assess the field's isotropy. But in other figures of the paper, where we compare
power spectra, azimuthal averages are used to make the plots easier to read.
It should also be noted that this Fourier-based method 
of generating fBms leads to periodic distributions and therefore our 
computations, unlike real observations, will not suffer from non-periodic boundary conditions\footnote{The reason why this is important is 
that numerical Fourier transform algorithms such as the FFT treat images as if they 
were periodic. If they are not, artificial jumps at the edges result in high frequency components which alter the power spectra. On the other hand, 
apodization of the images to prevent this may in turn affect large scales.}.

\subsection{3D density and velocity fields}

The purpose of this paper is to study the effects of projection from a three-dimensional dynamical medium onto a 
spectro-imagery observation. 
To conduct such a study, the first step is to generate model interstellar 3D density and velocity fields and from these
compute simulated PPV data cubes.

We therefore generated three-dimensional\footnote{The dimensions are 129 by 129 by 129 pixels.} fBms with spectral indexes 
of $-3.0$, $-3.5$, $-4.0$, and $-4.5$\footnote{These correspond to Hurst exponents respectively equal to 0, 0.25, 0.5, and 0.75.}. 
For each spectral index, 
two fBms were computed, one for the density field $n(x,y,z)$ and one for the velocity field 
$v(x,y,z)$. The computations were also 
performed with a uniform density field, for which the power spectral index is formally equal to $-\infty$.
{Here we just consider the line-of-sight (longitudinal) velocity component, 
as it is the only one available to observers. 
Like \cite{kitamura93} we make the assumption that the 3D velocity field $\vec{v}(x,y,z)$
is isotropic and homogeneous and that the longitudinal and lateral velocity components have the same power spectra
as $\vec{v}(x,y,z)$.}

For the sake of realism, some sort of correlation should exist between the 3D density and velocity fields. However, for simplicity, the main part of 
this paper deals with uncorrelated cubes, the effects of a certain amount of correlations between $n$ and $v$ being reported on in 
section~\ref{section_correl}.

\begin{figure}
\plotone{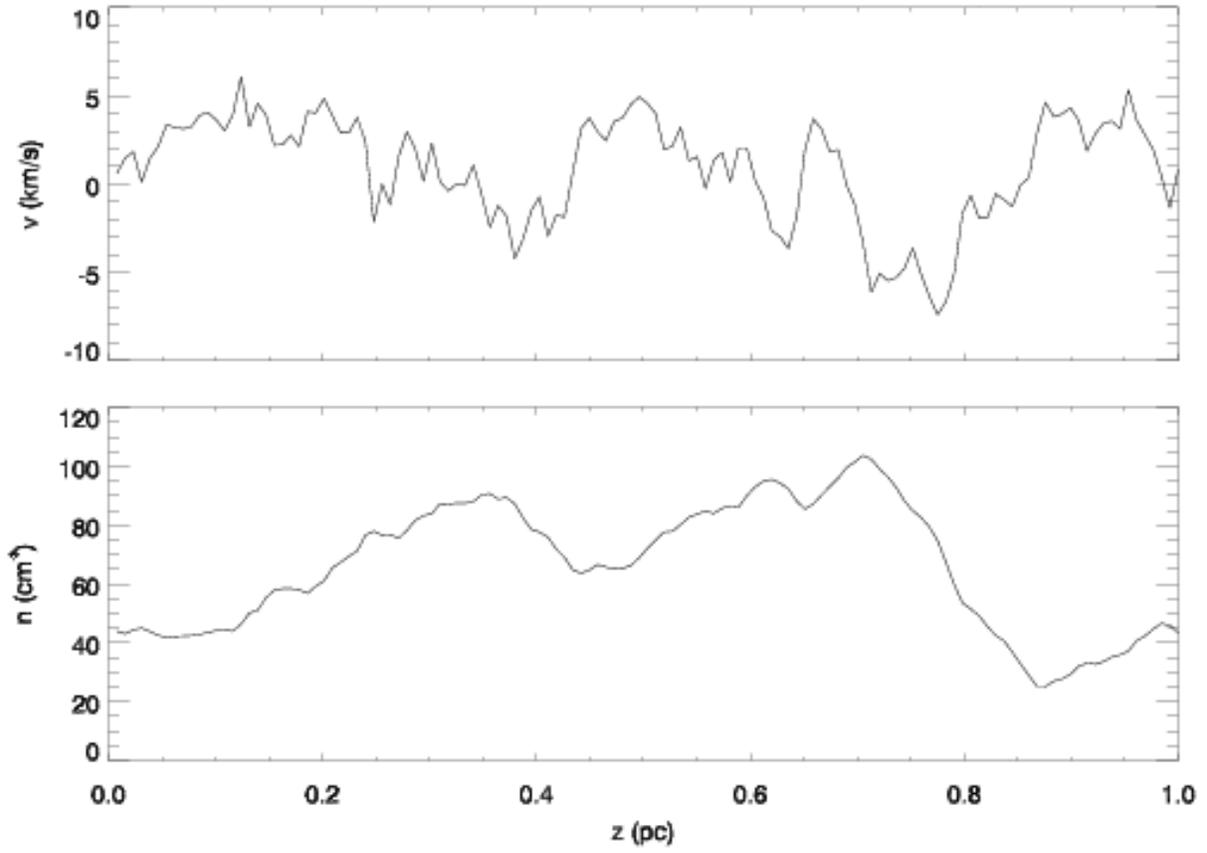}
\caption{\label{fig_los} Typical fractal line of sight extracted from the 3D density (bottom, $\gamma_n$=-4.5) and velocity (top, $\gamma_v$=-3.0) fields.}
\end{figure}

A typical line of sight extracted from one set of 3D density and velocity fields is shown in Fig.~\ref{fig_los}. The density and velocity fields used in 
this example have power spectral indexes of $\gamma_v=$-3.0 and $\gamma_n=$-4.5 respectively. 
The former displays more small-scale structures, as expected for a shallower 
spectral index $(\gamma_n < \gamma_v)$\footnote{\label{footnoteH}This criterion holds when comparing fields of the same Euclidean dimension, but the relevant number
 in direct space when assessing the amount of small-scale structures is the Hurst exponent, as will be discussed later. Here $\gamma_n < \gamma_v$ 
is equivalent to $H_n > H_v$.}. By construction, the mean values of the original cubes are set by the normalization $A_0$ and their variances are 
determined by the total number of pixels. To obtain physically plausible numbers, the following procedure was applied:
The minimum values of the density cubes were subtracted from them in order to obtain only positive values, and a multiplicative factor was applied to 
obtain an average density of $90~\mathrm{cm}^{-3}$. For the velocity cubes, we first subtracted the mean, then scaled the values to obtain a 
velocity dispersion of 3~\kmsp.
An important feature to point out is that, since typically only $0.1\%$ of the points in a Gaussian 
distribution are more than $3\sigma$ away from the mean, our density fields present quite low contrasts, with $\sigma/{\bar{n}}\lesssim 0.3$. 
This is an unavoidable limitation of these model fields. 

\section{Simulation of a PPV data cube} 

\label{sec_simul_ppv}

\subsection{Definition}

To obtain a PPV data cube from the simulated 3D density and velocity fields, we compute, at each position $(x,y)$, the spectrum that would be 
observed in this direction. This is done under the simplifying assumption that the medium is observed in an 
optically thin line, such as, in some instances, the 21 cm 
line of neutral atomic hydrogen. In this case, the spectrum at a given position $(x,y)$ is computed by simply adding the contributions of all cells 
along the line of sight, the emission from a single gas cell being a 
Gaussian line centered at velocity $v(x,y,z)$ and of 
velocity-integrated area proportional to $n(x,y,z) \delta z$
\footnote{Here $\delta z$ is the depth, in cm, of a cell in the 3D density cube. The total depth of the cloud was set to 1 pc and the size of the fBm is $129 \times 129 \times 129$. In this case, $\delta z=2.414 \times 10^{16}$ cm}. 
Eventually, the PPV data cube $N_u(x,y,u)$, which is the column density of {\sc Hi} along the line of sight $(x,y)$ at the velocity $u$ within $\delta u$, can be expressed by the following equation\footnote{Here we consider the particular case where the spectral resolution of the observation is equal to the channel width and where 
the spectral transmission is a step function of width $\delta u$.}:
\begin{equation}
\label{eq_PPV}
N_u(x,y,u)=\sum_z \frac{n(x,y,z)\delta z}{\sqrt{2\pi}\sigma(x,y,z)} \exp \left ( -\frac{[u-v(x,y,z)]^2}{2\sigma(x,y,z)^2} \right ).
\end{equation}
The dispersion of the Gaussian is given by:
\begin{equation}
\sigma(x,y,z)=\sqrt{ \left[\frac{\partial v(x,y,z)}{\partial z} \delta z\right]^2 +  \frac{k_B T}{m} },
\end{equation}
where $k_B$ is the Boltzmann constant, 
$T$ is the gas kinetic temperature and $m$ is the mass of the emitting species (in our case {\sc Hi}).  
This broadening includes not only the thermal velocity dispersion $\sigma_{th}=\sqrt{k_BT/m}$, but also a small-scale velocity
gradient which roughly models the subpixel structure of the velocity field \citep{brunt2002a}. Also, to make sure that we lose a 
negligible fraction of the signal, the velocity limits in the final PPV cube are given by $v_{\mathrm{min}}-2\sigma_{th}$ and $v_{\mathrm{max}}+2\sigma_{th}$, 
where $v_{\mathrm{min}}$ and $v_{\mathrm{max}}$ are the minimum and maximum values in the 3D velocity field. The number of channels is then 
obtained by dividing this velocity range by the channel width $\delta u$, which was set to $0.25~\mathrm{km}~\mathrm{s}^{-1}$. As a result, all of our PPV cubes do not have the same extent in the third dimension, since it depends on the 3D velocity 
cube and on the temperature.

\subsection{Synthesized spectra}

\begin{figure}
\plotone{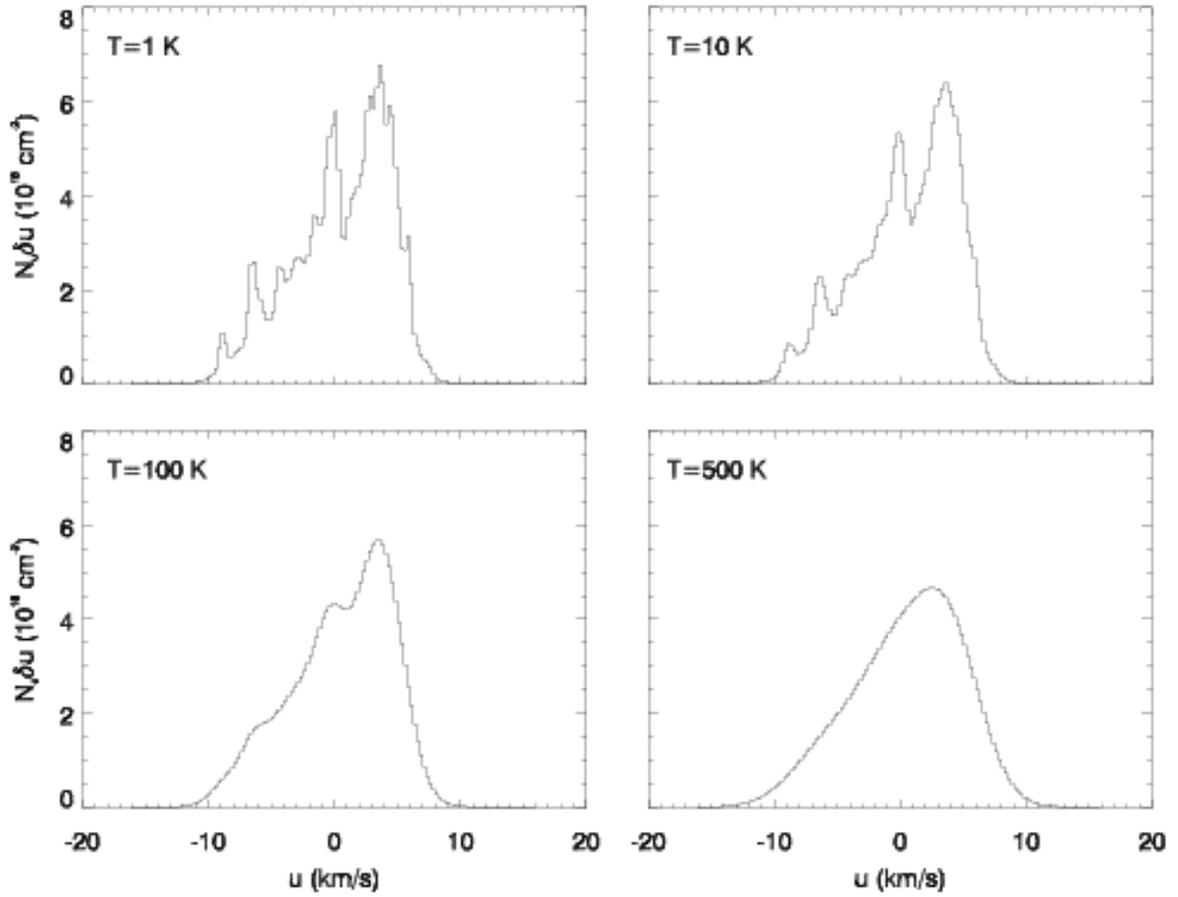}
\caption{\label{fig_example_spectra} Synthetic spectra for the line of sight described in Fig.~\ref{fig_los}, at kinetic temperatures $T=1$~K, 10~K, 100~K 
and 500~K. The spectral resolution was set to $\delta u = 0.25~\mathrm{km}~\mathrm{s}^{-1}$.}
\end{figure}

Examples of the line synthesis procedure described above are shown in Fig.~\ref{fig_example_spectra} for the same line of sight that 
was used in Fig.~\ref{fig_los}. These spectra were computed using equations~\ref{eq_PPV} for $T=$1~K, 10~K, 100~K, 
and 500~K.
One can see the expected smearing effect of the temperature, 
which smooths the spectra as $T$ increases, similar to the effect of decreasing the spectral resolution. 
As the temperature increases, the spectrum becomes smoother 
and closer to Gaussianity, indicating that the thermal broadening is close to
the original dispersion in the input 3D velocity cube. 
For reference, the thermal dispersions for all temperatures are 
$0.09~\mathrm{km}~\mathrm{s}^{-1}$ at $T=$1~K, $0.29~\mathrm{km}~\mathrm{s}^{-1}$ at 
$T=$10~K, $0.91~\mathrm{km}~\mathrm{s}^{-1}$ at $T=$100~K
and $2.0~\mathrm{km}~\mathrm{s}^{-1}$ at $T=$ 500~K.

The spectrum computed at $T=$ 100~K, a value representative 
of {\sc Hi} gas, is reminiscent of what is observed at 21 cm in high latitude clouds. 
At artificially low temperature\footnote{In {\sc Hi}, temperatures seem to vary from around 80~K in 
the cold neutral medium (CNM) to 5000~K - 8000~K in the warm neutral medium (WNM).} ($T=$ 1~K) many components appear. These components do not correspond to any physical ``cloud'' structure, as they simply represent 
the fractal properties of velocity and density fluctuations along the line of sight. 
At such a low kinetic temperature, the line is far from the optically thin regime and the effect of opacity on the observed spectrum would be important. These spectra are shown only 
to illustrate the effect of kinetic temperature on our ability to resolve kinematical structures. 
The effective spectral resolution of such PPV observations therefore includes the
thermal broadening. Considering, as a simplification, a Gaussian-shaped channel of full width at half-maximum (FWHM) $\delta u$, this effective resolution is approximately
\begin{equation}
\label{eq_delta_veff}
\delta v_{eff} \approx \sqrt{\delta u^2 + (2.16)\sigma_{th}^2 }.
\end{equation}

\subsection{Channel maps}

\begin{figure}
\plotone{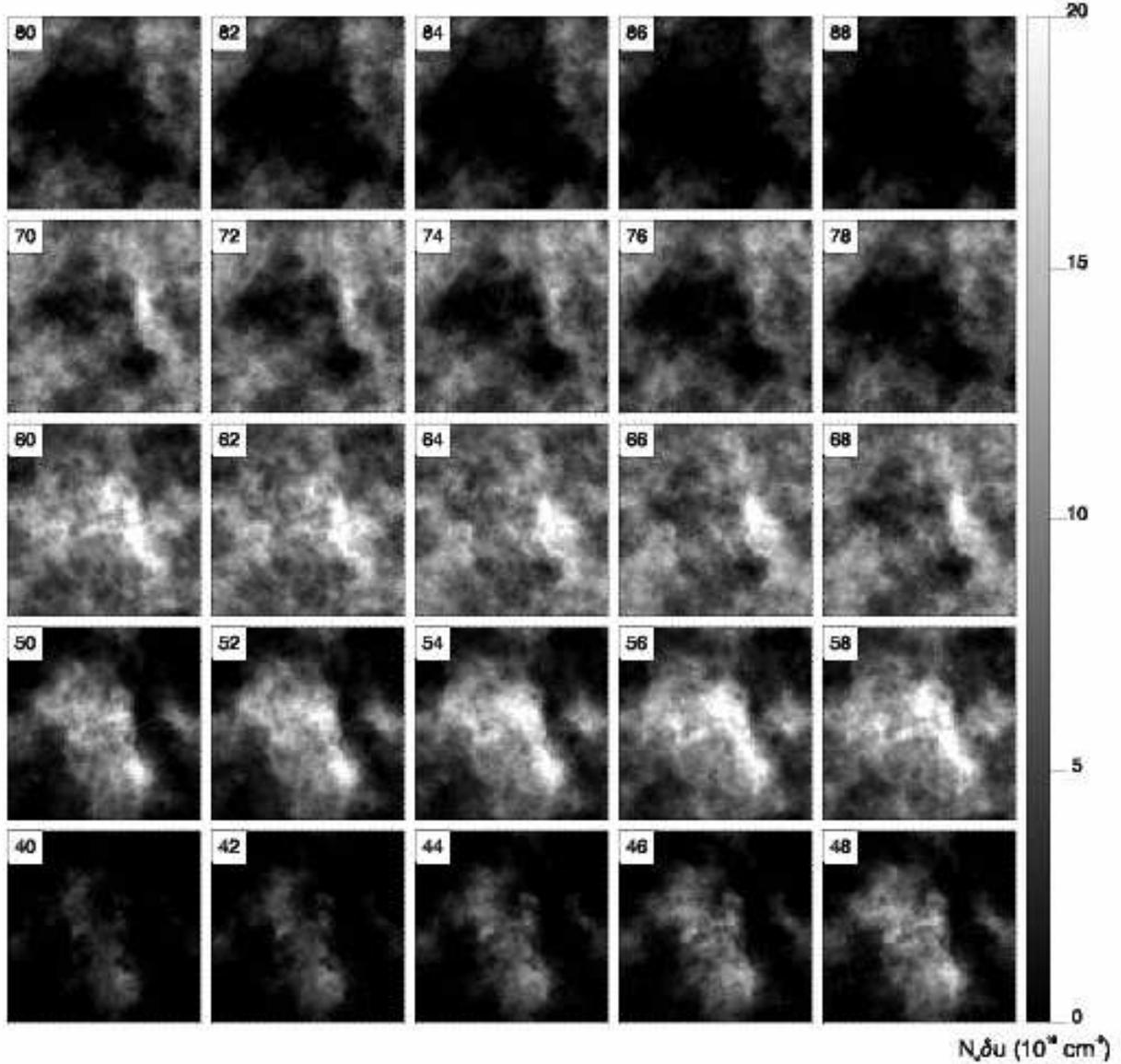}
\caption{\label{fig_example_cm} Selected channel maps from a typical simulated PPV data cube, shown here for comparison with actual observations.
In this case the power spectrum index of the density and velocity fields are respectively $\gamma_n=-3.5$ and 
$\gamma_v=-3.0$. The gas temperature used is $T=100$~K (thermal velocity dispersion $\sigma_{th}=0.91~\mathrm{km}~\mathrm{s}^{-1}$) and the 
spectral resolution is $\delta u=0.25~\mathrm{km}~\mathrm{s}^{-1}$. 
The channel number is indicated in the upper left corner of every channel map.}
\end{figure}

Selected channel maps from the PPV cube computed with $\gamma_n=-3.5$ and $\gamma_v=-3.0$ are shown in Fig.~\ref{fig_example_cm}. 
These channel maps are reminiscent of what is observed at 21 cm for instance \citep[]{joncas92}. 
The structures in neighbouring channel maps are clearly 
correlated, with both elongated and diffuse structures. If it is partly due to the channel width being subthermal, correlation still exists between more widely separated channels. 
\section{Statistical properties}

\label{sec_stat_prop}

\begin{figure}
\plotone{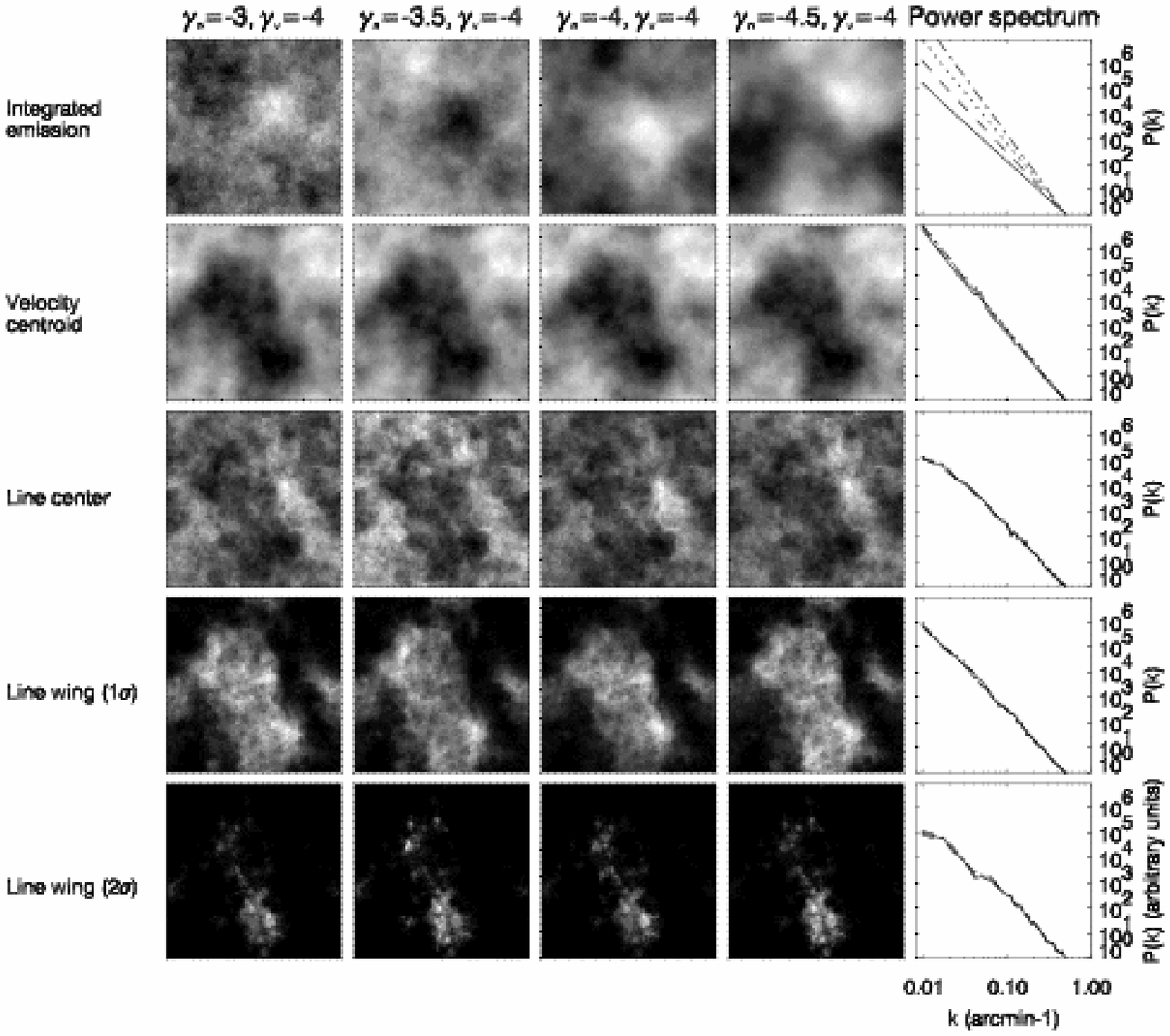}
\caption{\label{fig_channels_v4} Integrated emission, centroid velocity and three velocity channels for 
$\gamma_v=-4$ simulations. The four image columns correspond to four values of $\gamma_n$ specified above each column. 
The power spectra of all 2D maps shown here are displayed on the far right column, the solid, dashed, dotted and dash-dotted lines 
representing respectively $\gamma_n=-3.0$, -3.5, -4.0, and -4.5. 
This figure highlights the fact that the power spectrum of the integrated emission is completely determined by
the 3D density field (we verify that their power spectrum index are equal).
But the most striking feature of this figure is the fact that the centroid velocity is independent of the
density field and that the power spectrum of the centroid velocity is equal to the power spectrum of 3D velocity field.
In addition, the structure of channel maps and their power spectra are mostly determined by the velocity field
(we observe a slight variation of the channel maps structure with different density fields).}
\end{figure}

\begin{figure}
\plotone{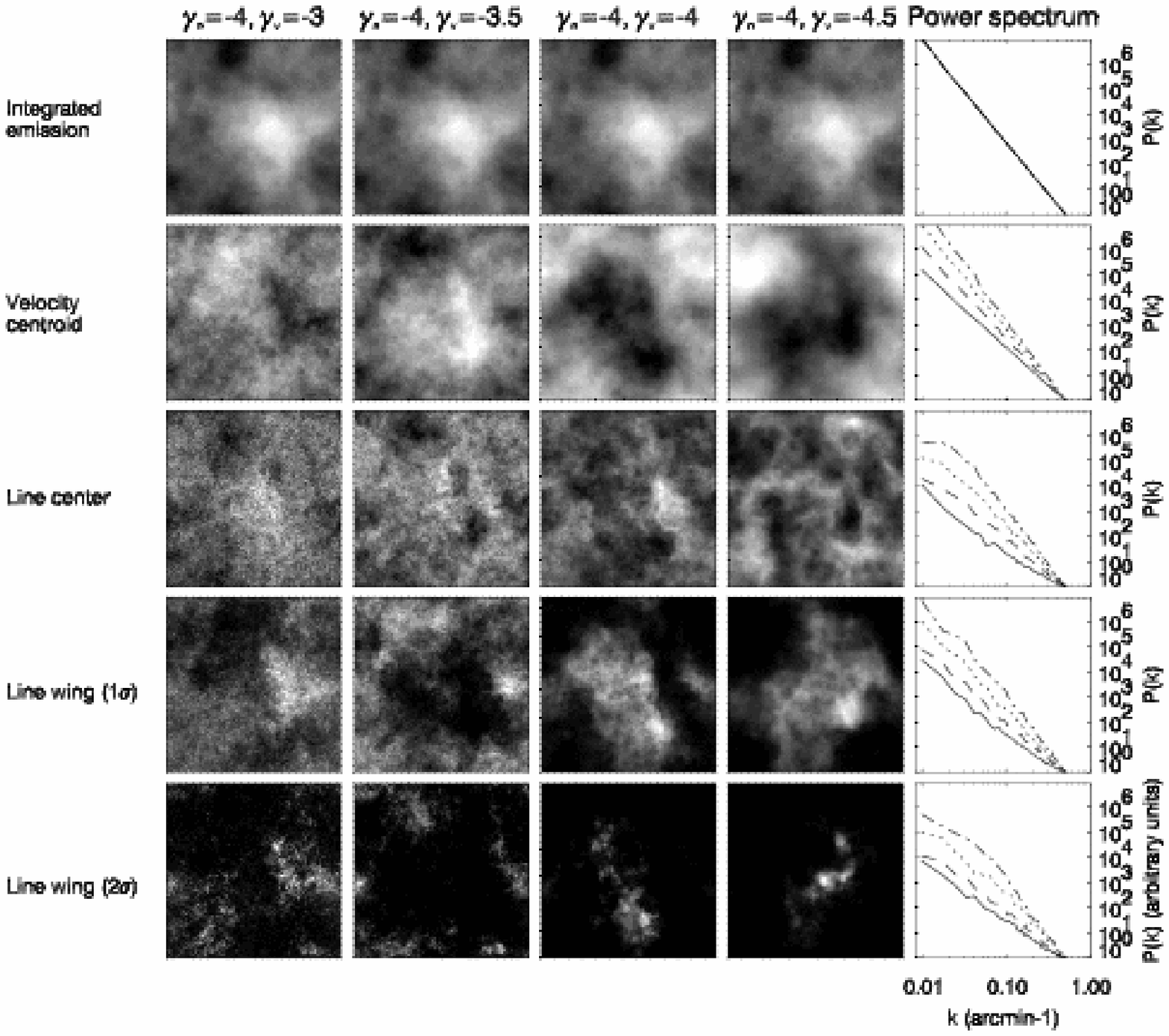}
\caption{\label{fig_channels_n4} Integrated emission, centroid velocity and three velocity channels for 
 $\gamma_n=-4$ simulations. The four image columns correspond to four values of $\gamma_v$ specified above each column. 
The power spectra of all 2D maps shown here are displayed on the far right column, the solid, dashed, dotted and dash-dotted lines 
representing respectively $\gamma_v=-3.0$, -3.5, -4.0, and -4.5. This figure confirms the fact that the integrated emission
is completely independent of the velocity field (top row) and that the power spectra of the velocity centroid and channel maps
are completely dominated by the 3D velocity field. This figure also shows the effect of ``shot-noise'' discussed in the text
that produces a flattening at small scales of the channel maps power spectrum, for shallow ($\gamma_v=-3$) velocity fields.}
\end{figure}

In this section the power spectra of individual channel, integrated emission and centroid velocity maps, obtained from the 
PPV data cubes, are examined and linked to the statistical properties of the 3D density and velocity fields. To conduct our analysis, PPV cubes with all possible combinations of 3D density and velocity fields were simulated. Fig.~\ref{fig_channels_v4}
and \ref{fig_channels_n4} display the integrated emission map, 
the centroid velocity map, 3 selected channel maps - one in the line center and two in the line wings - and their associated
power spectra, for a subset of data cubes. More precisely, Fig.~\ref{fig_channels_v4} shows the variations of these quantities for $\gamma_v=-4$ and
all values of $\gamma_n$, while Fig.~\ref{fig_channels_n4} presents the same analysis for $\gamma_n=-4$ and
all values of $\gamma_v$. A detailed analysis of these figures follows.

\subsection{Integrated emission}

\label{section_int_emission}

It has been shown by several authors (see for instance \citet{stutzki98} and \citet{goldman2000}) 
that, for optically thin gas with Gaussian statistics, the power 
spectral index of the integrated emission map is exactly the power spectrum of the 3D density field, under the condition $d/L \geq 1$ where
$L$ is the largest scale on the $(x,y)$ plane and $d$ is the depth in the $z$ dimension.
A brief summary of the derivation of this result is given in the appendix. This is true 
whether the power spectrum is a power law or not, provided isotropy is maintained. 
This property of optically thin media is powerful as it allows to directly deduce the power 
spectrum of the interstellar 3D density field from the observed map.

\begin{figure}
\plotone{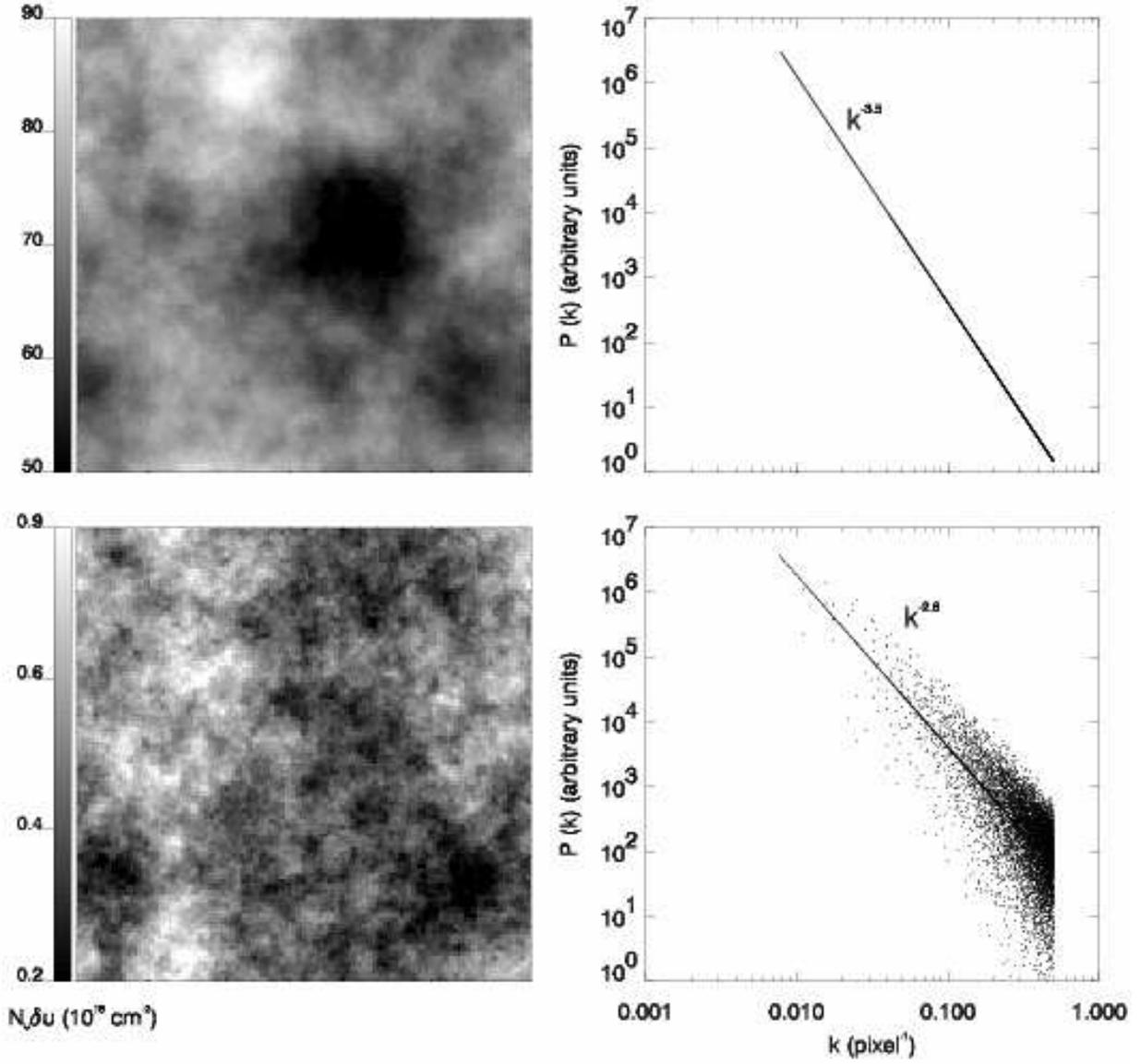}
\caption{\label{fig_coldens_vs_slice} Map and power spectra of the fully integrated emission (top row) and of a single slice (bottom row) 
through one of the 3D fBm density cubes, in this case $\gamma_n=-3.5$ and $\delta u=0.25$ \kmsp.}
\end{figure}

Our simulations confirm this property for the PPV cube integrated emission 
(see the top panels of Fig.~\ref{fig_channels_v4} and \ref{fig_channels_n4})
and for the image of the 3D density field integrated along the $z$ axis (see Fig.~\ref{fig_coldens_vs_slice}).
The fact that the 3D density field and the integrated emission - which is
its projection onto the plane of the sky - should have the same spectral
index does not mean, however, that they exhibit the same amount of structure as a
function of scale. As already suggested in footnote~\ref{footnoteH}, the relevant number
in this comparison is the Hurst exponent. Now, since $\gamma_{\mathrm{3D}}=-2H_{\mathrm{3D}}-3$ and
$\gamma_{proj}=-2H_{proj}-2$, the preservation of the spectral index means that
$H_{proj}=H_{\mathrm{3D}}+0.5$, an effect called projection smoothing.

On the other hand a cut through the 3D density field would statistically represent the original
field, and the Hurst exponent would remain the same $H_{cut}=H_{\mathrm{3D}}$. The power spectrum of such a cut
would then have an index $\gamma_{cut}=-2H_{cut}-2=\gamma_{\mathrm{3D}}+1$. 
An example of this is shown in Fig.~\ref{fig_coldens_vs_slice} for a 1-pixel slice of the 
3D density cube \footnote{In this figure, the power spectrum shows a fair amount of dispersion and the index is a bit lower than $\gamma_{\mathrm{3D}}+1$. This can be understood by the fact that our slice through the 3D density cube is not a ``perfect'' one, but indeed has a certain width. The dispersion effect is more pronounced at small scales, which are comparable to this width.}.

\begin{figure}
\plotone{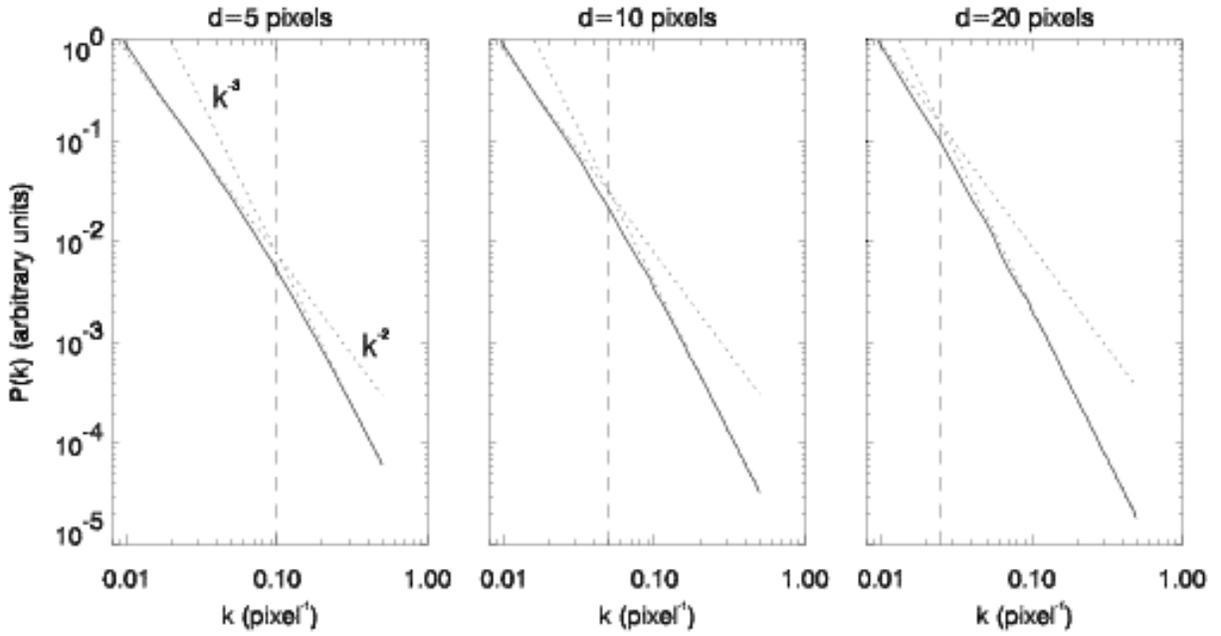}
\caption{\label{fig_ps_vs_d} Power spectra of integrated emission slices of a $\gamma_n=-3.0$ density cube.
The depth of the slices are $d=5$, 10 and 20 pixels, from left to right. The solid line is the power spectrum
of the integrated density slice. The dotted lines indicate $-3.0$ and $-2.0$ slopes and the 
vertical dashed line shows the $1/2d$ scale. }
\end{figure}

It is also interesting to study the power spectrum of a slice in the intermediate regime, 
where $0<d<L$. To study this, the power spectra of density field slices of 
increasing thickness $d$ were computed. Fig.~\ref{fig_ps_vs_d} presents the power spectra of three cases
for $d$=5, 10 and 20 pixels. The 3D density field used here has $\gamma_n=-3$.
The power spectra are bent, with a spectral index of $\gamma_n$ at small scales and $\gamma_n+1$ at large scales. 
The transition occurs at a frequency $k_d = 1/2d$ and provides a method to determine the depth of the observed medium.
{This is in accordance with the pioneer work of \citet{von_hoerner51} (summarized by \citet{odell87})
who studied the effect of projection smoothing on the index of the structure function.}
Recently, \citet{elmegreen2001} used this property to determine the thickness of the \hi layer in the Large Magellanic Cloud.
The absence of such a curvature in the power spectra of large scale \hi Galactic emission 
suggests in turn that the depth of the \hi layer sampled by the observations is significantly larger than
the map size, i.e. much larger than 25 pc in the case of the high latitude cloud observed by \cite{miville-deschenes2003a}.
This result is also relevant for studies of optically thick media (e.g. molecular clouds) where the observed emission
comes only from the surface of the cloud. In that case, the spectral index measured on the integrated
emission map of an optically thick tracer should be closer to $\gamma_n+1$.

\subsection{Centroid velocity}

Another quantity built from PPV data cubes is the centroid velocity map, which, for any line of sight, is simply an average velocity weighted by the line temperatures, 
\begin{equation}
\label{eq_centroide1}
C(x,y)=\frac{\sum_u u~N_u(x,y,u) \delta u}{\sum_u N_u(x,y,u) \delta u}.
\end{equation}

This equation is easily shown to be equivalent to the density-averaged 
velocity component parallel to the line of sight \citep{dickman85}:
\begin{equation}
\label{eq_centroide2}
C(x,y)=\frac{\sum_z v(x,y,z)~n(x,y,z) \delta z}{\sum_z n(x,y,z) \delta z}.
\end{equation}

Centroid velocity maps have traditionally been used to study gas kinematics in various ISM environments.
For instance in the study of the turbulent energy cascade in \hii regions \citep{miville-deschenes95}
and in molecular clouds \citep{miesch94}. But one of the main difficulties of these analyses
is to relate the statistical properties of the centroid velocity map to the 3D velocity field. 

To assess this relation in the framework of our fBm simulations, we computed the centroid velocity maps from PPV cubes for every combination of $\gamma_n$ and $\gamma_v$. Some of them are shown in Fig.~\ref{fig_channels_v4} and Fig.~\ref{fig_channels_n4}, along with their power spectra, and all measured spectral indexes are given in Table~\ref{tab_centroid_exponent}. In the range of density and velocity indexes tested, the spectral index $\gamma_C$ of the centroid velocity map is remarkably equal to $\gamma_v$, whatever the power spectrum of the density field. Fig.~\ref{fig_channels_v4} illustrates this behaviour: the velocity centroid maps computed for the different values of $\gamma_n$ are the same, and their power spectra all merge into the same 
spectrum. 

\begin{figure}
\plotone{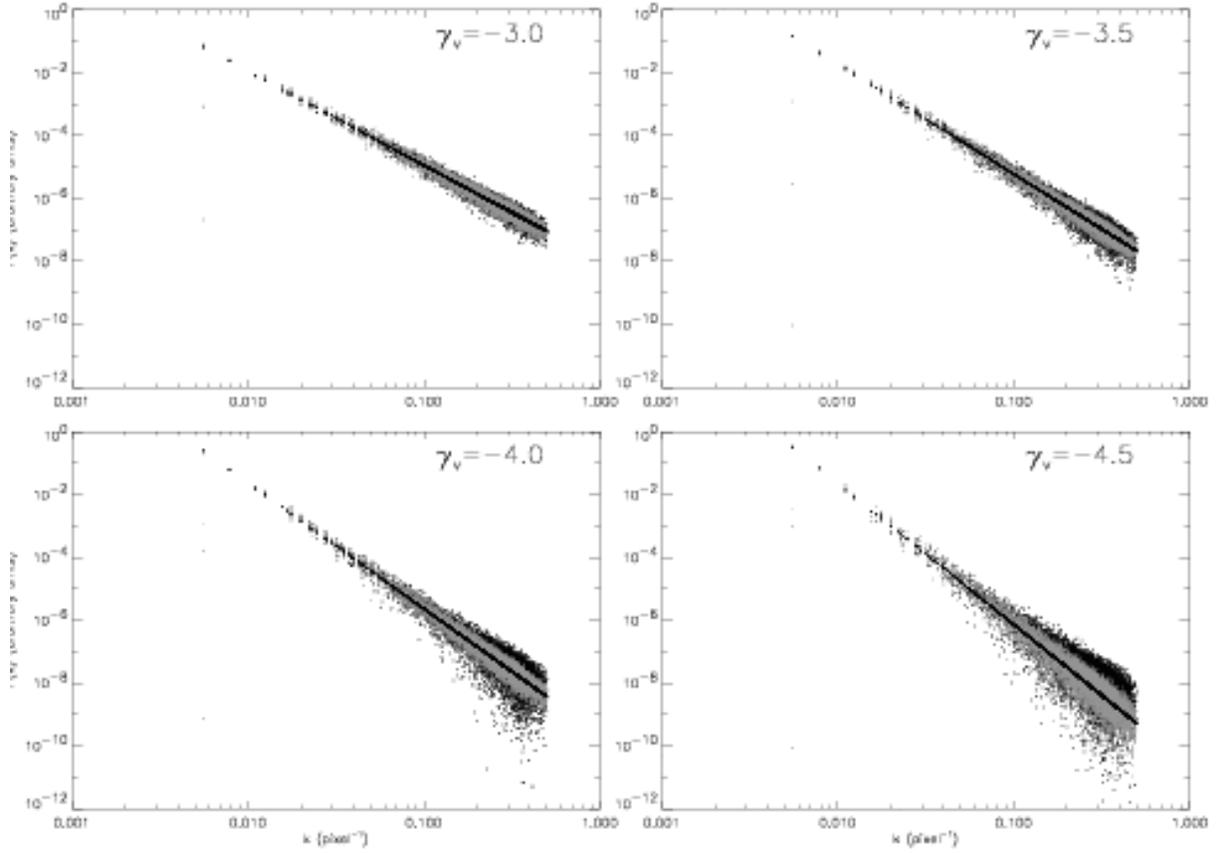}
\caption{\label{fig_centroide_dispersion} Power spectra of centroid velocity maps for various ($\gamma_n$, $\gamma_v$) combinations. Each subplot corresponds to a given $\gamma_v$ and features the cases for $\gamma_n = -3.0$ (black scatter plot), $\gamma_n = -4.0$ (grey scatter plot), and uniform density (thin black line). The latter is actually a scatter plot with virtually zero dispersion, as expected.}
\end{figure}

This result is easy to obtain analytically for a uniform density, as is shown in the Appendix (see also \citet{dickman85}). 
But, to our knowledge, this is the first time that it is demonstrated for stochastic density fields. 
Fig.~\ref{fig_centroide_dispersion} displays the full scatter plots of power versus wavenumber for several ($\gamma_n$, $\gamma_v$) combinations. It appears that the dispersion around the power law decreases with steepening density spectra, and in the limit of a uniform density field, isotropy is recovered. Conversely, for a given $\gamma_n$, the centroid spectra are more scattered for steeper values of $\gamma_v$. This may be explained by the fact that density fluctuations manifest themselves more strongly with smoother velocity fields.

\begin{figure}
\plotone{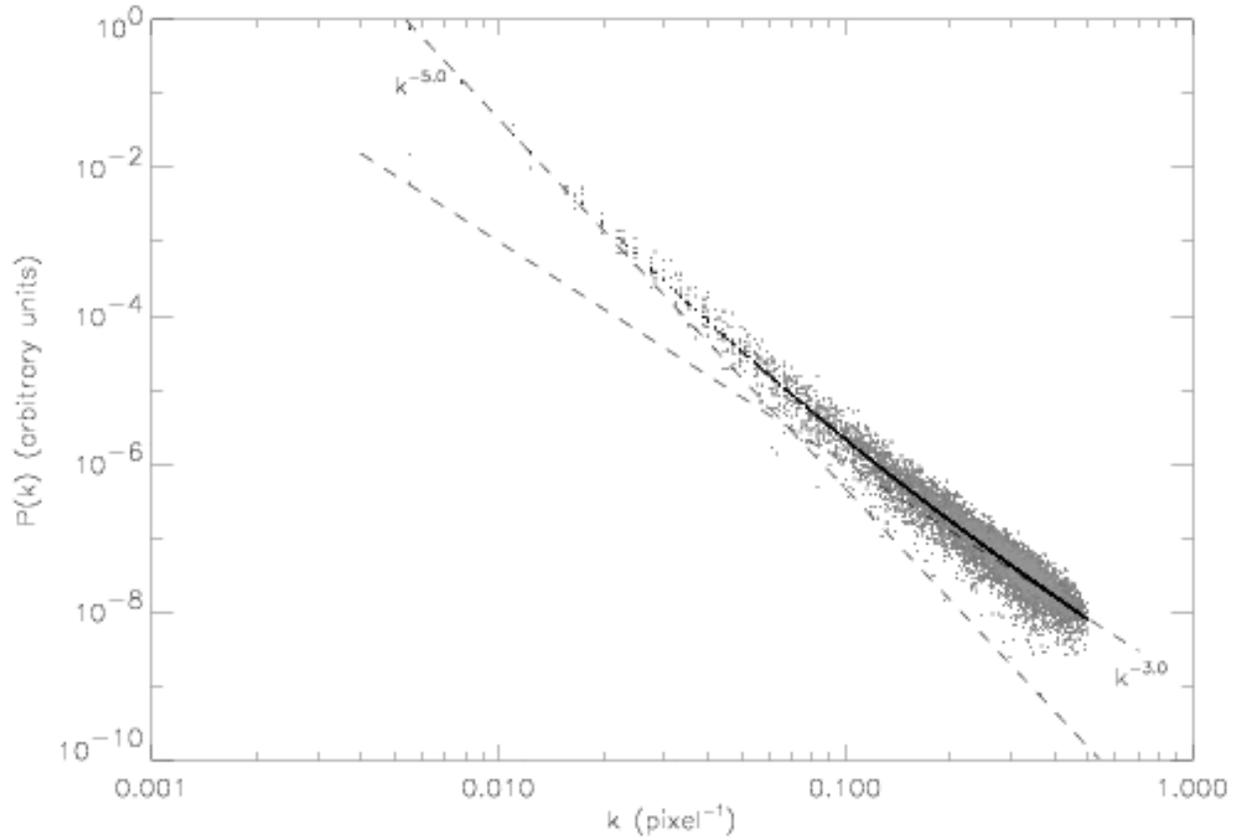}
\caption{\label{fig_bent_centroide} Power spectra of centroid velocity maps constructed from a velocity field with a curved power spectrum (see text). Plots for the uniform density case (black) and the $\gamma_n=-4$ case are presented, along with the expected asymptotic behaviours at small ($k^{-3}$) and large ($k^{-5}$) scales.}
\end{figure}

Further tests of this result have been performed. For instance, in accordance with Eq.~\ref{eq_centroide2}, it does not depend on the gas kinetic temperature, with $\gamma_C$ being equal to $\gamma_v$ in all cases considered. More interesting was the test of non-fBm density and velocity fields. One of these involved a velocity field for which the power spectrum was bent, varying from a $k^{-5}$ behaviour at large scales (small $k$) to a $k^{-3}$ behaviour at small scales (large $k$). For this, two fBms with respective indexes -3.0 and -5.0 were built and their Fourier amplitudes added with proper weights\footnote{This is to ensure that the velocity field is not dominated by either behaviour over the whole range of $k$, in which case the curvature of the power spectrum would be imossible to see.}. After introduction of a random phase map, an inverse Fourier transform was performed. The velocity field computed this way was then used with two density cubes, one with $\gamma_n=-4.0$ and the other being the uniform density cube, to simulate two PPV data sets. Velocity centroids for both cases were then computed and their power spectra are shown in Fig.~\ref{fig_bent_centroide}. For the uniform density case, unsurprisingly, the power spectrum of the velocity centroid has zero dispersion and exhibits exactly the same variation as that of the full 3D velocity field. For the $\gamma_n=-4.0$ case, despite the dispersion, the behaviour remains, showing the consistency of the method, even for isotropic velocity fields with non-power-law power spectra.

\begin{figure}
\plotone{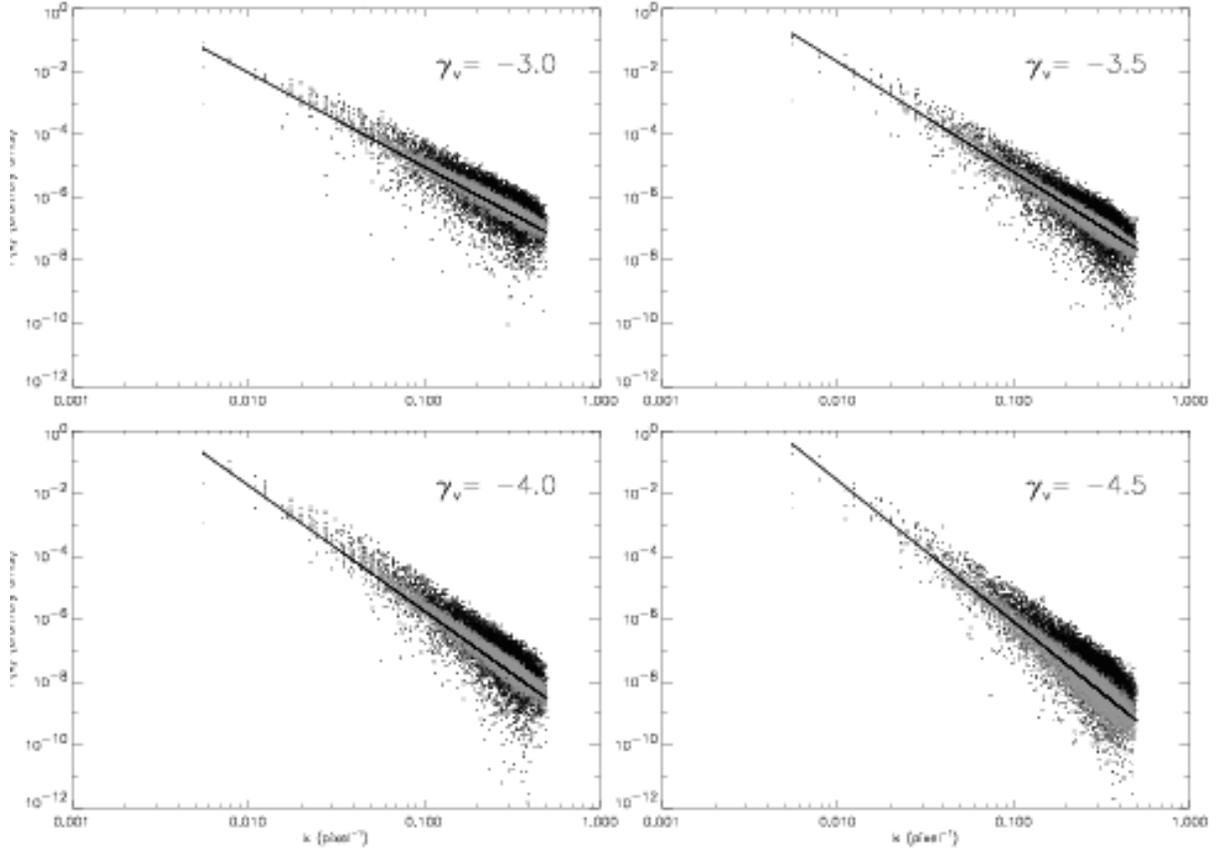}
\caption{\label{fig_centroide_lognormal} Power spectra of centroid velocity maps constructed from a lognormally distributed density field (see text). Each subplot corresponds to a given $\gamma_v$ and features the lognormal density case (black scatter plot), and the fBm density case with $\gamma_n=-4$(grey scatter plot). The black line represents the expected behaviour in $k^{-\gamma_v}$.}
\end{figure}

As a second extension of the types of fields tested, density cubes with increased fluctuations were built by exponentiation of fBm fields, resulting in lognormal distributions. The original fBm field used had an index $\gamma=-4.0$ and the resulting density values ranged from $\sim$~1~cm$^{-3}$ to $\sim$~2500~cm$^{-3}$, with a mean of $\sim$~90~cm$^{-3}$. This density cube was used in conjunction with all of our fBm velocity fields to compute PPV data sets and velocity centroids, the power spectra of which are shown in Fig.~\ref{fig_centroide_lognormal}. For such large density fluctuations, the dispersion is noticeably increased with respect to the fBm case, but the main result that $\gamma_C=\gamma_v$ holds when averaging in wavevector space.

Considering the possibility that the complete agreement between $\gamma_C$ and $\gamma_v$, without $\gamma_n$ entering into the picture, 
could be caused by the absence of density-velocity correlations, we chose to compute 3D density and velocity fBm fields featuring a certain 
level of correlation. The way we did this is based on the idea that most, if not all, of the spatial structure of images is contained in the structure 
of their Fourier phases. Starting from this, we introduced correlations between two fBms by setting a number of phase points to the same value 
in both fields. The points were chosen randomly, and their number varied between 10\% and 90\% of the total number of pixels. The amplitudes 
were computed separately. 

Once again, we found that the velocity centroid spectral index was exactly the same as that of the three-dimensional 
velocity field, whatever the density spectral index, and whatever the amount of correlations. From this remarkable result we infer that 
velocity centroids are robust observables for recovering 3D velocity statistics from optically thin lines.
This result has obvious applications for studies of ISM statistical properties and of interstellar turbulence 
(see \citet{miville-deschenes2003a}).

{We are aware that the method used to introduce density-velocity correlation in our simulations does not comply with (magneto-)hydrodynamics, as for instance high densities would correspond to high velocities along the line of sight, but this technique allows us to test easily various correlation degrees while staying in the fBm scheme. Obviously, it would be useful to make a similar analysis on physically relevant fields, such as MHD simulations, to study the effects on the centroid velocity power spectrum of physical density-velocity correlation, sharper density/velocity contrasts and non-Gaussian statistics, and thus assess the robustness of this result in more realistic conditions.}

One should also keep in mind that our results may not be generally applicable to all interstellar conditions.
For instance the statistical properties of centroid velocity maps can be significantly affected by systematic 
motions like rotation or expansion, if proven not to be part of the turbulent cascade itself. In this case, 
such systematic velocity fields should be filtered prior to any analysis of turbulent motions \citep{kleiner85}.

\label{section_correl}

\subsection{Channel maps and velocity slices}

In a recent study, LP00 showed that, for Gaussian random fields, the spectral index of the 3D velocity field can be 
determined by studying the variation of the spectral index of velocity slices\footnote{A velocity slice of thickness $\Delta V$ is the 
integrated emission of the PPV cube between $v$ and $v+\Delta V$.} as a function of their thickness $\Delta V$.
Subsequently, \citet{lazarian2001} and \citet{esquivel2003} analyzed and confirmed this proposition using MHD simulations. 
{In both these works the power spectrum of the simulated velocity field 
is forced to be a power law (a method called spectral modification) to extend the limited power law range of MHD simulations. 
This method preserves the phase information and thus much of the expensively-obtained physics. }
We propose here to also test the theoretical predictions of LP00 on fBm simulations, because they have 
{intrinsically} well defined power spectra and obey Gaussian statistics, the working hypotheses of LP00.

\subsubsection{Thickness of velocity slices}
 
The main prediction of LP00 deals with what they call {\it thin} and
{\it thick} velocity slices of the PPV data cube. These properties are
scale-dependent: A velocity slice of width $\Delta V$ is considered thin at a scale $l$ if the
velocity dispersion in the observed medium at that scale $\sigma(l)$ is larger than $\Delta V$. Conversely, a velocity slice is said to be 
thick at scale $l$ if $\sigma(l) < \Delta V$. Since the velocity dispersion depends on $l$ according to 
\begin{equation}
\label{eq_sig_l}
\sigma(l) = \sigma(L) \left( \frac{l}{L} \right)^{(-\gamma_v-3)/2.},
\end{equation}
where $\sigma(L)$ is the total velocity dispersion\footnote{$L$ is the largest scale in the image.} (set to 3~\kms in our simulations), velocity slices are thinner at large scales than at small scales ($\gamma_v < -3$). 
In fact, from Eq.~\ref{eq_sig_l}, the smallest 
scale $l_{thin}$ for which a velocity slice of width $\Delta V$ can be considered thin is given by:
\begin{equation}
\label{eq_l_thin}
l_{thin} = L \times \left( \frac{\Delta V}{\sigma(L)} \right)^{\frac{2}{-\gamma_v-3}}.
\end{equation}
Given the values of $\Delta V$ and $\sigma(L)$ sampled in our simulations, the velocity slices under study can be thin
or thick at all scales, in which cases they are respectively said to be {\it fully thin} or
{\it fully thick}. They can otherwise exhibit a transition between the two regimes, being thin at large scales and thick at small scales. Obviously, a velocity slice will be fully thin if all sampled scales 
are larger than $l_{thin}$, and fully thick if the largest scale in the map is smaller than $l_{thin}$.

At that point it is important to check whether or not our simulations allow us to investigate the thin regime. The velocity dispersion at
the 1 pixel scale (see Table~\ref{tab_thinthick}) is given by Eq.~\ref{eq_sig_l}, and is to be compared with the smallest $\Delta V$ available, which is determined by the effective spectral resolution (see Eq.~\ref{eq_delta_veff}), namely $\delta v_{eff}=1.4$~\kms for $\delta u=0.25$~\kms and $T=100$~K. 
It appears that, for $\gamma_v=-3$, there are meaningful velocity slices (with $\Delta V \geq \delta v_{eff}$) in the fully thin regime, since $\sigma(l=$1 pixel) $> \delta v_{eff}$. For smaller values of $\gamma_v$, however, any velocity slice will be thick at small scales. Nonetheless, the values of $l_{thin}$, given for all $\gamma_v$ in Table~\ref{tab_thinthick},
show that a $\Delta V=\delta v_{eff}$ velocity slice is
thin on a large range of scales, even for $\gamma_v=-4.5$. 
In order to have access to fully thin velocity slices for all $\gamma_v$,
an increased dispersion $\sigma(L)$ or a lower temperature would have been necessary (see Eq.~\ref{eq_sig_l}), but that would have led to more velocity channels in the PPV cube\footnote{The spectral resolution $\delta u$ being kept constant, in order to keep $\delta v_{eff}$ unchanged.} and therefore to more shot-noise, which has 
a dramatic effect on simulated channel maps, as will be discussed later.

The condition expressed above for a velocity slice of width $\Delta V$ to be fully thin calls for a remark. The criterion $\sigma(l) > \Delta V$ is given for two-dimensional scales $l$ in the plane of the sky, but the velocity dispersion of the gas sampled at that scale by the observations is actually a function of the 3D scales $l_{3\textrm{D}}$. Many of those, when projected, give the same 2D scale $l$, but their velocity dispersions can be quite different, and an average should enter in the thinness condition. Its present form however ensures that, since $l_{3\textrm{D}} \geq l$, if the velocity slice is thin at scale $l$, it will also be thin at all unprojected scales $l_{3\textrm{D}}$, according to Eq.~\ref{eq_sig_l}.

According to LP00, fully thin velocity slices, when integrated, produce maps whose spectral index is (for $\gamma_n < -3$)
\begin{equation}
\label{eq_laz1}
\gamma_{thin}=-3-\frac{\gamma_v+3}{2},
\end{equation}
and fully thick velocity slices exhibit a spectral index 
\begin{equation}
\label{eq_laz2}
\gamma_{thick}=-3+\frac{\gamma_v+3}{2}.
\end{equation}
When a substantial part of the emission is integrated over, the slices are
said to be {\it very thick} and their spectral index tends to that of the 3D density field,
as shown in \ref{section_int_emission}.

\begin{figure}
\plotone{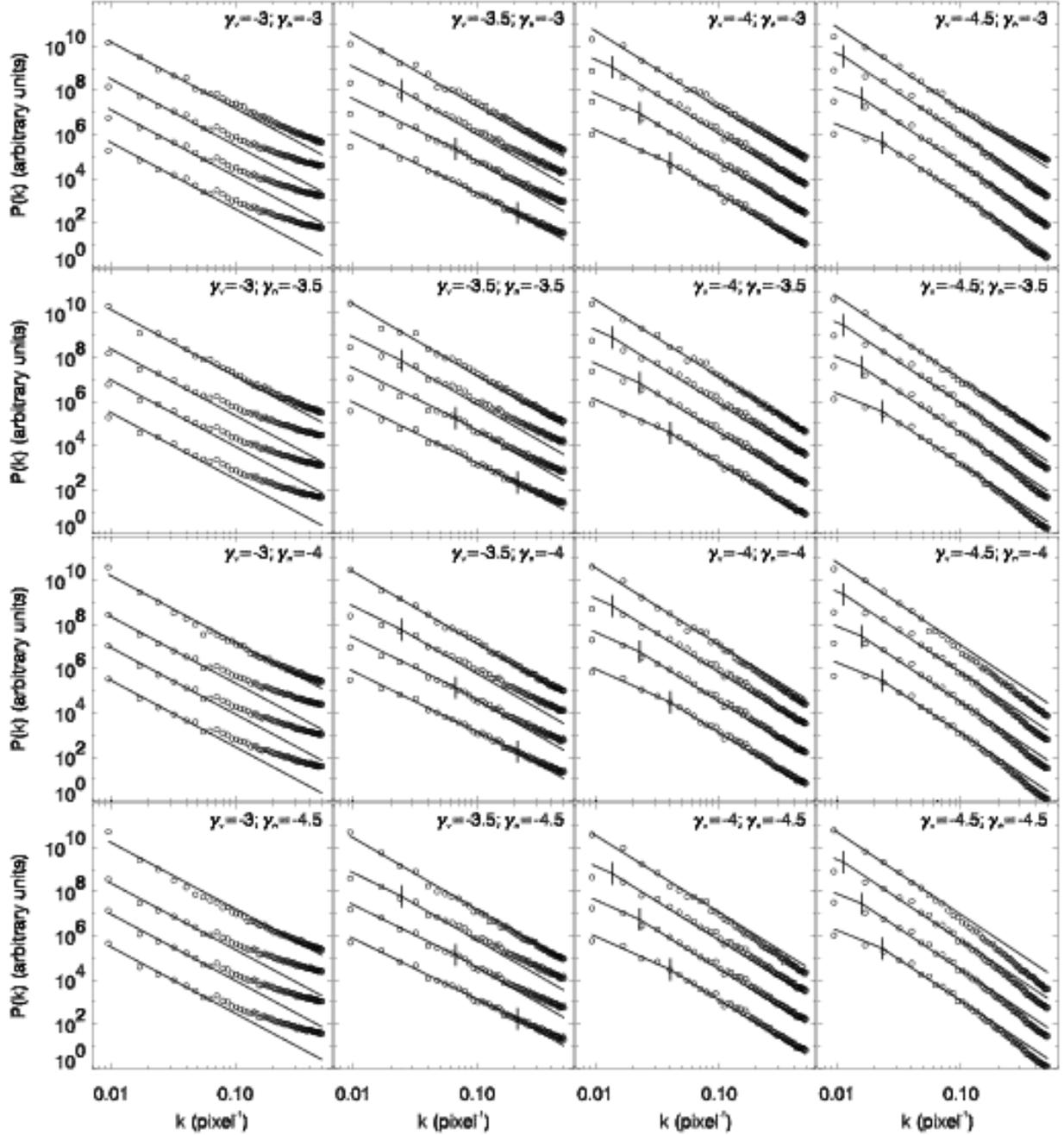}
\caption{\label{fig_ps_deltav_allgamma} Power spectra of velocity slices for all ($\gamma_n$, $\gamma_v$) combinations. 
On each plot, the open circles represent the power spectra of velocity slices for $\Delta V=1.25,$ 1.75, 2.25 and 10.0 \kms 
(from bottom to top). The solid line on each curve is the prediction of LP00 (see equations~\ref{eq_laz1} and \ref{eq_laz2}), normalized
to fit around the thin-thick transition (indicated by a vertical marker). On each curves, the scales are thin (thick) on
the left (right) side of this vertical line.}
\end{figure}

To test this prediction, the spectral indexes of velocity slices of 
increasing width were computed, for all our PPV data cubes, and at a temperature of $T=$100~K. 
The power spectra of selected velocity slices with different widths $\Delta V$ are shown in Fig.~\ref{fig_ps_deltav_allgamma}. Each plot corresponds to a different set of ($\gamma_n$, $\gamma_v$) and shows, for the given PPV cube, the power spectra
of four velocity slices (open circles) with $\Delta V=1.25$, 1.75, 2.25 and 10.0 \kmsp (from bottom to top). 
In each case, the solid lines represent the predictions of LP00 for the thin and thick regimes (see Equations~\ref{eq_laz1}
and \ref{eq_laz2}), and the small vertical marker indicates the transition scale $l_{thin}$, given
by Eq.~\ref{eq_l_thin}. The velocity slice under consideration is thin for scales on the left of this mark, and thick for those on the right.
The power spectra shown in Fig.~\ref{fig_ps_deltav_allgamma}  
are in general agreement with the predictions of LP00, given the selected values
of $\Delta V$, $\gamma_n$ and $\gamma_v$.
Bent power spectra are observed, with a break close to $l_{thin}$. However, the narrow range of scales over which the velocity slices are thin makes the comparison with the predictions uneasy. Furthermore, the power spectra of velocity slices for $\gamma_v=-4.5$ move away from the LP00 prediction 
at small scale, because these velocity slices are becoming very thick at those scales. This behavior is indeed correlated with $\gamma_n$.
Another feature of Fig.~\ref{fig_ps_deltav_allgamma} is the increase of power at small scales for $\gamma_v=-3$, due to shot-noise.

\subsubsection{Finite-size or ``shot-noise'' effect}

\label{sec_shot_noise}

The computation of PPV cubes from finite size 3D velocity fields leads to an artificial 
increase of high-frequency fluctuations in channel maps \citep{lazarian2001,brunt2002a,esquivel2003}. As pointed out by \citet{esquivel2003},
the limited size of the simulated lines of sight introduces sharp edges and almost empty channels
in the PPV cube. This effect is more important with a small effective spectral resolution.
Such a high-frequency flattening is seen in the power spectra of channel maps for 
$\gamma_v=-3$ in figures~\ref{fig_channels_n4} and \ref{fig_ps_deltav_allgamma}. 
As expected, the amplitude of this high-frequency component increases when the size of the 3D fields decreases. This effect biases the determination of simulated 
channel maps power spectra. 

\begin{figure}
\plotone{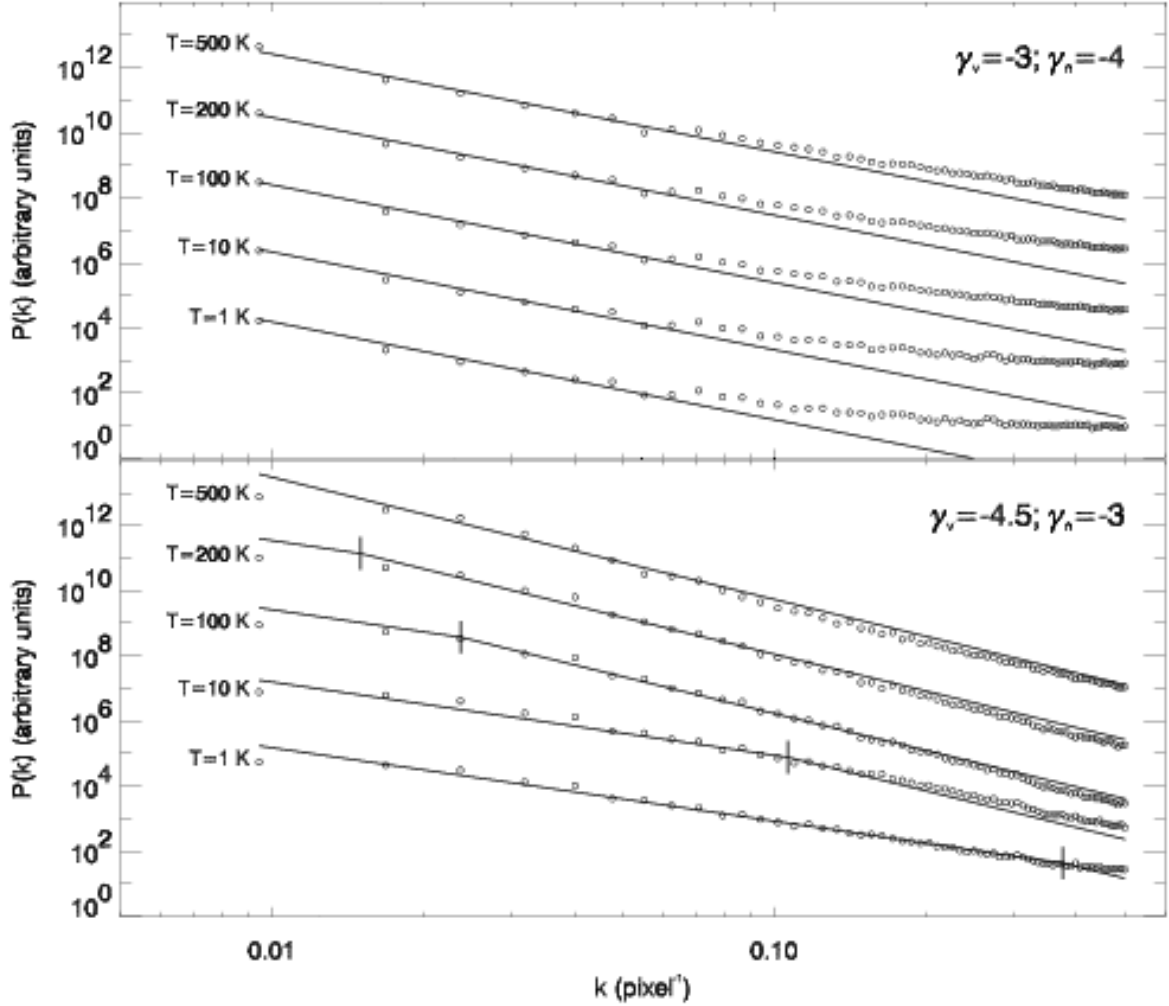}
\caption{\label{fig_shot_noise} Power spectra of 1 pixel velocity slices of width $\Delta V=1.25$~\kms
for ($\gamma_n=-4,$ $\gamma_v=-3$) (top) and ($\gamma_n=-3,$ $\gamma_v=-4.5$) (bottom). 
On each plot, the five curves (open circles) represent PPV cubes computed with $T=1$, 10, 100, 200 and 500 K.
The solid line on each curve is the prediction of LP00 (see equations~\ref{eq_laz1} and \ref{eq_laz2}), normalized
to fit around the thin-thick transition (indicated by a vertical marker). The slices are thin for scales on the left side of these vertical lines, and thick for scales on their right side.}
\end{figure}

One solution to reduce the effects of shot-noise is to apply a Gaussian smoothing to the 
PPV spectra, simulating the effect of thermal broadening. But such a smoothing
also increases the effective spectral resolution (see Eq.~\ref{eq_delta_veff}) which
results in a reduction of the accessible thin regime.
An example of this is shown in Fig.~\ref{fig_shot_noise} where we plot the power spectra
of velocity slices with $\Delta V=\delta_{veff}$, for two PPV cubes 
(($\gamma_n=-4$, $\gamma_v=-3$) and ($\gamma_n=-3$, $\gamma_v=-4.5$))
computed with different temperatures ($T=1$, 10, 100, 200 and 500 K).
The $\gamma_v=-3$ case is particularly interesting as velocity slices
are thin at all scales as long as $\Delta V \leq \sigma(L)$ (see Eq.~\ref{eq_sig_l}), which is the
case here ($\sigma(L)=3$ \kms and $\delta v_{eff} = 2.9$ \kms for the highest temperature $T=500$ K).
This allows for a separate study of the effect of Gaussian smoothing on shot-noise reduction and velocity slice thinness.
The top panel of Fig.~\ref{fig_shot_noise} shows that Gaussian smoothing significantly dampens shot-noise. Up to a temperature of 500 K the power spectrum steepens through reduction of the shot-noise effect, but it seems that, even at this temperature, shot-noise has not been completely removed. At higher temperatures, the condition $\delta v_{eff} \leq \sigma(L)$ would not be met any longer.

The bottom panel of Fig.~\ref{fig_shot_noise} illustrates the effect of Gaussian smoothing
on both the shot-noise and the thickness of the velocity slice. At low temperature ($T=$1 K),
the velocity slice is thin on almost all scales but the effect of shot-noise can be observed at high-frequency
as the power spectrum is above the prediction of LP00. As expected, the shot-noise amplitude 
and the range of scales at which the velocity slice is thin both decrease with increasing $T$. At high temperature, the shot-noise is completely
removed but the velocity slice is fully thick.
This shows how the use of Gaussian smoothing is necessarily a compromise between
the reduction of shot-noise and the thinness of velocity slices. 
From the power spectra shown in Fig.~\ref{fig_ps_deltav_allgamma}, shot-noise
appears to be significant only for $\gamma_v=-3$, indicating that the simulation parameters 
selected ($\sigma(L)=3$~\kmsp, $T=100$ K and $\delta u=0.25$~\kmsp)
provide a good compromise, for most of our $\gamma_v$ values, between shot-noise reduction 
and channel map thinness.

\subsubsection{Velocity slice analysis}

\begin{figure}
\plotone{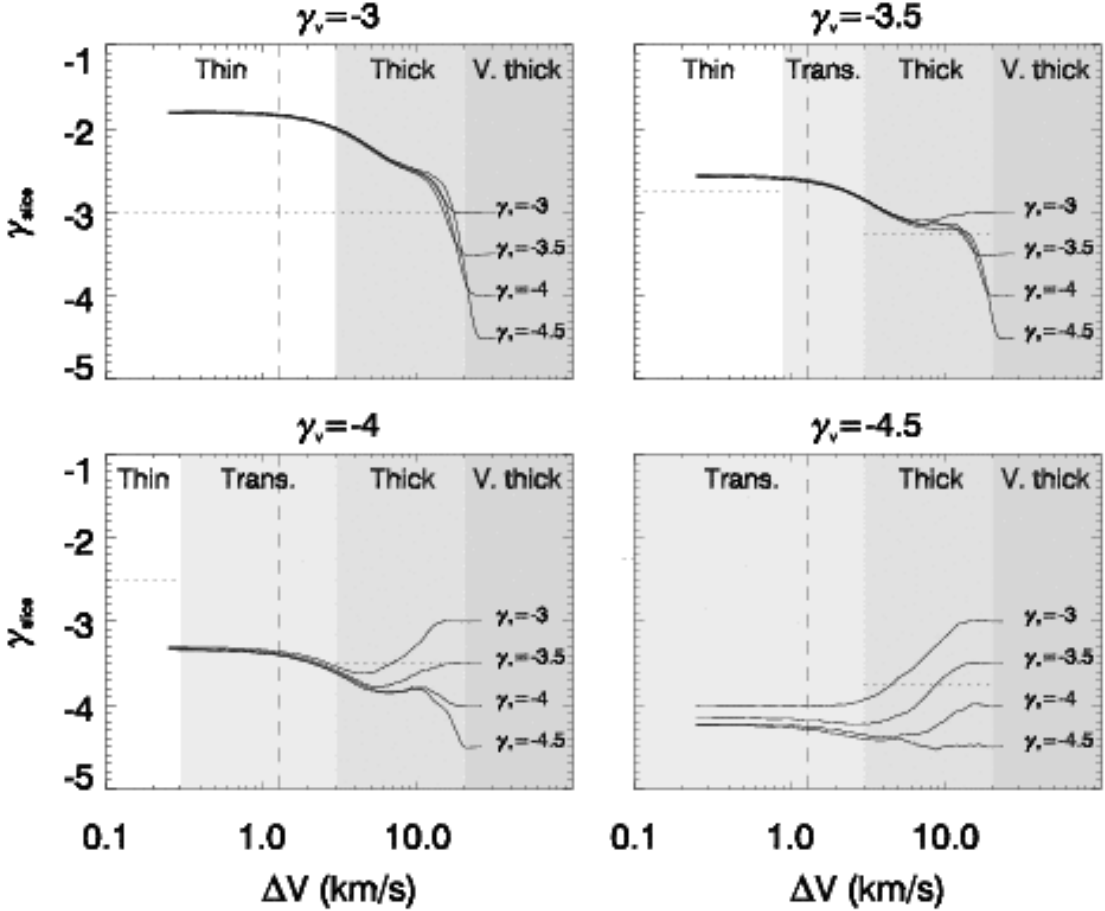}
\caption{\label{fig_thinthick} Velocity slice spectral indexes as a function of velocity width $\Delta V$,
computed for all ($\gamma_n$, $\gamma_v$) combinations. Curves are grouped by their $\gamma_v$ value. 
The white and grey zones indicate the ``thin'', ``transition'', ``thick'' and ``very thick''
regimes. The thin regime is for $\Delta V < \sigma(l=1$ pixel). The thick regime is for
$\Delta V > \sigma(L)$ (the velocity width is larger than the velocity dispersion at the largest scale).
The transition regime corresponds to velocity slices with intermediate $\Delta V$ (they are thin at large scales 
and thick at small scales). The very thick regime is where $\gamma_{slice}=\gamma_n$ ($\Delta V \sim 20$~\kmsp).
The vertical dashed line indicates the effective spectral resolution of our simulations ($\delta_{veff}=1.4$~\kmsp).
The horizontal dotted lines indicate LP00's predictions for $\gamma_{thin}$ and $\gamma_{thick}$ (see equations~\ref{eq_laz1}
and \ref{eq_laz2}).}
\end{figure}

The method proposed by LP00 to determine $\gamma_n$ and $\gamma_v$ is to look at the spectral
index ($\gamma_{slice}$) of velocity slices of increasing width $\Delta V$. According to LP00, the curve of $\gamma_{slice}$ vs $\Delta V$ should reach two asymptotic values at small (thin) and large (very thick) $\Delta V$ 
that allow the determination of $\gamma_n$ and $\gamma_v$. 
This velocity slice analysis has been performed on our simulations, and the results are presented in Fig.~\ref{fig_thinthick},
which shows the variations of the index $\gamma_{slice}$ 
(fitted over the whole range of $k$) as a function of the velocity width $\Delta V$.
The curves, one for every set of ($\gamma_n$, $\gamma_v$), are grouped by their $\gamma_v$ value. 
The shaded areas in this figure indicate the fully thin, fully thick and very thick regimes
but also the transition regime where the velocity slice is thin at large scales and
thick at small scales. 
The vertical dashed line indicates the effective spectral resolution of our simulations. 
The plateau seen on the left of this line reflects the fact that velocity structures cannot be
resolved due to the finite spectral resolution.

As $\Delta V$ enters the very thick regime, the spectral index $\gamma_{slice}$ converges 
to the $\gamma_n$ value, in agreement
with the fact that the integrated emission has the same power spectrum as the 3D density field.
At the other end, we confirm the prediction of LP00 that the power spectrum of narrow velocity slices 
is independent of the density field. 
For a given $\gamma_v$ all curves converge to the same value for small $\Delta V$, except in the case of $\gamma_v=-4.5$, as explained in the next paragraph.
This strong correlation between the 3D velocity field and channel maps is also seen in Fig.~\ref{fig_channels_v4}
and \ref{fig_channels_n4}. First, in Fig.~\ref{fig_channels_v4}, the channel maps' structure
is dominated by the velocity field as it varies very little with $\gamma_n$. Similarly, the channel maps power spectra for a given $\gamma_v$ are the same for all values of $\gamma_n$.
The influence of the velocity field on the channel maps structure is also manifest in Fig.~\ref{fig_channels_n4}
where channel maps power spectra vary with $\gamma_v$. 

Back to Fig.~\ref{fig_thinthick}, the curves for $\gamma_v=-3$ rise at small $\Delta V$, up to a value of -1.8.
As the spectral index $\gamma_{slice}$ has been computed over the whole $k$ range, this trend 
can be completely attributed to shot-noise, which has a significant effect 
for $\gamma_v=-3$ (see Fig.~\ref{fig_shot_noise}).
For $\gamma_v=-3.5$ the effect of shot-noise is reduced, and in that case, the curves are in relatively good agreement with the predictions of LP00. There is a significant
plateau in the thick regime, at a value very close to the predicted $\gamma_{thick}$ value (see Eq.~\ref{eq_laz2}).
Moreover, for $\Delta V=\delta_{veff}$, we are very close to the fully thin regime and the spectral index 
measured there is consistent with LP00.
For $\gamma_v=-4$ and -4.5, the concordance with LP00 is not that clear. In these two cases the effect of shot-noise
is completely removed by the Gaussian smoothing but, on the other hand, we are very far from the fully thin 
velocity slices. Some kind of plateau is observed in both cases within the thick regime, but not at the predicted $\gamma_{thick}$ value. This is due to the fact that the smallest scales are actually becoming very thick. This is why, in the case of $\gamma_v=-4.5$, the curves are dependent on $\gamma_n$ and do not converge to the same value.

\subsubsection{Discussion}

Except for $\gamma_v=-3.0$, we have seen that shot-noise has a limited effect on our simulations
because of the Gaussian smoothing applied. But this is at the expense of a narrowing of the accessible thin 
regime: fully thin velocity slices with $\Delta V \geq \delta v_{eff}$ are present only when $\gamma_v=-3.0$, 
In these conditions, the interpretation of the $\gamma_{slice}$ vs $\Delta V$ curves 
(see figures~\ref{fig_shot_noise} and \ref{fig_thinthick}) is ambiguous, which makes
the determination of $\gamma_v$ very difficult. The power spectra of fully thick velocity slices can actually depend on $\gamma_n$, which 
makes it impossible to deduce $\gamma_v$ from $\gamma_{thick}$ using Eq.~\ref{eq_laz2}.
{Therefore, due to the relative thickness of velocity slices in our simulations, we could not demonstrate whether the LP00 method can be used to determine $\gamma_v$.}

One practical limitation of the LP00 method, however, is that it requires to find out whether or not the fully thin regime is reached, for a given observation. 
If a significant asymptotic plateau is found at small $\Delta V$ 
in the $\gamma_{slice}$ vs $\Delta V$ curves, one might think that it is the case.
But unfortunately, as seen in Fig.~\ref{fig_thinthick}, such a plateau could also be
observed in the thick regime, and this could lead to a wrong estimate of $\gamma_v$.
In addition, the only way to estimate the position of the transition scale $l_{thin}$ between the two regimes for a given velocity slice is by using
Eq.~\ref{eq_l_thin}, which requires $\gamma_v$. Therefore, there is no independent
way to determine whether thin velocity slices have been reached or not.
In that respect, the determination of $\gamma_v$ in the Small Magellanic Cloud 
by \citet{stanimirovic2001} might be questionable as they do not seem to reach 
the thin asymptotic regime.

{It is important to estimate in what conditions LP00's method 
could be used on real observations, for a typical high latitude \hi cloud for instance.
We have shown that the narrowest relevant velocity slice for this analysis
has a width of $\Delta V = \delta_{veff} \sim\sqrt{2.16\, k_B T/m}$ (here we consider 
that the effective spectral resolution is dominated by the thermal broadening).
But to estimate $\gamma_v$ using LP00's method one must have a plateau at small $\Delta V$ in the 
$\sigma_{slice}$ vs $\Delta V$ curve, which calls for fully thin velocity slices 
on a significant range of $\Delta V$, lets say for $\delta_{veff} \leq \Delta V \leq 2 \, \delta_{veff}$.
We recall that velocity slices are thin at scales $l \geq l_{thin}$ where $\sigma(l_{thin}) = \Delta V$.
Therefore to evaluate on what scales LP00's method could be used to deduce $\gamma_v$, 
one must assess what is $l_{thin}$ for a velocity slice of $\Delta V = 2 \, \delta_{veff}$.
As shown by \citet{miville-deschenes2003a}, the energy spectrum of \hi is compatible 
with the \citet{kolmogorov41} cascade ($\sigma(l)= A\, l^{0.33}$).
In these conditions, $l_{thin}$ for a velocity slice
of $\Delta V= 2 \, \delta_{veff}$ is
\begin{equation}
l_{thin} = \left(\frac{8.64 \, k_B T}{m A^2}\right)^{3/2}.
\end{equation}
The normalisation factor $A$ can be estimated from high-latitude \hi observations
like in \citet{joncas92}, where the internal velocity dispersion at a scale of 1 pc is of the order
of 4 \kmsp, significantly larger than that observed in molecular clouds 
\citep{larson81}. If we consider  Cold Neutral Medium (CNM) gas at T=100 K, the smallest thin 
scale $l_{thin}$ for a cloud at a distance of 100 pc is $\sim 8'$,
which is significantly larger than the angular resolution (1') 
of the 21 cm radio-interferometer at Dominion Radio Astrophysical Observatory (DRAO).
Furthermore, if we consider the presence of warmer gas  (Warm Neutral Medium ($T=5000$-8000 K)
or thermally unstable \hi ($T\sim 1000$ K)), the thermal broadening  would increase $l_{thin}$ 
(which scales as $T^{3/2}$) and make the determination of $\gamma_v$ using the LP00 method 
only possible at scales larger than a few degrees, at least for \hi in the solar neighborhood. 
In this respect, the velocity centroid method is better suited to determine the 3D velocity power spectrum as it does not depend at all
on the gas temperature and on the thickness of the velocity slices in the observation. Therefore the velocity centroid method
allows the determination of the velocity power spectrum down to the angular resolution of the observations.}

\section{Conclusion}

\label{sec_conclu}

In this paper three-dimensional fBms with various spectral indexes were used to simulate interstellar 3D density and velocity fields. 
From these we computed several spectro-imagery observations assuming optically thin line emission. 
Our results confirm that the spectral index $\gamma_W$ of the integrated emission is the same as that of the 3D density field $\gamma_n$, when 
the depth $d$ of the medium probed along the line of sight is at least of the order of the largest transverse scale observed. For smaller depths, the power spectrum
of the integrated emission shows two asymptotic slopes of $\gamma_n$ and $\gamma_n+1$ at small and large scales respectively, with the transition occurring at a frequency $k=1/2d$. This property of the power spectrum of integrated emission could be used to
estimate the depth of interstellar clouds and to infer the 3D statistical properties of optically thick media.

The main result of our study is that the
spectral index of the maps of line centroid velocities is the same as that
of the 3D velocity field. This result is particularly robust because it holds for any
gas temperature and whether density-velocity correlations are present or
not. This similarity is also recovered if the power spectrum of the
3D velocity field is not a power law or if the density fluctuations follow a lognormal distribution.
This remarkable property of optically thin media may be used to estimate 
the 3D energy spectrum of interstellar turbulence in various ISM environments.

We also compare the power spectra of velocity slices with the predictions of LP00.
{Due to the limited resolution of our simulations we are not 
able to confirm their predictions.}
On the other hand we show that the method proposed by LP00 is in practice barely applicable to real observations, 
and we suggest that a more robust determination of 3D velocity statistics lies with velocity centroids.

\appendix

\section{Spectral index of the integrated emission}

In our case where the medium under study is optically thin, the integrated emission map is simply the sum of all velocity channels:
\begin{equation}
W(x,y)=\sum_u N_u(x,y,u) \delta u.
\end{equation}
Using Eq.~\ref{eq_PPV}, reverting the order of summations over $z$ and $u$, and with our values of $\delta z$ and $\delta u$, 
this translates into
\begin{equation}
W(x,y)=\alpha \sum_z n(x,y,z)\delta z
\end{equation}
where $\alpha$ is a numerical coefficient coming from the unit conversion. The important point is that $W$ is proportional to the sum 
over $z$ of the 3D density cube, which by construction is a fBm with spectral index $\gamma_n$. To compute the spectral index 
$\gamma_W$ of the integrated emission, we consider the limit for which the summation can be replaced by an integral over $z$
\begin{equation}
W(x,y)=\alpha \int n(x,y,z)\mathrm{d}z.
\end{equation}
Writing the Fourier transform of the 3D density as a function of the three-dimensional wave vector ${\bf k}$, we have
\begin{equation}
\tilde{n}({\bf k}) = \int\!\!\!\int\!\!\!\int n({\bf r})e^{-2i\pi {\bf k}.{\bf r}}\mathrm{d}x\mathrm{d}y\mathrm{d}z.
\end{equation}
Which becomes, if we consider the $k_z=0$ cut through $\tilde{n}({\bf k})$ and integrate first over $z$, 
\begin{equation}
\tilde{n}(k_x,k_y,0) = \int\!\!\!\int \frac{W(x,y)}{\alpha}e^{-2i\pi (k_xx+k_yy)}\mathrm{d}x\mathrm{d}y
\end{equation}
meaning that the Fourier transform of $W$ is proportional to the $k_z=0$ cut through $\tilde{n}$. We therefore have
\begin{equation}
|\tilde{W}(k_x,k_y)|^2~\propto~|\tilde{n}(k_x,k_y,0)|^2~\propto~(k_x^2+k_y^2)^{\gamma_n/2}, 
\end{equation}
and it is then easy to see that $P_W$ 
and $P_n$ follow the same law, and thus
\begin{equation}
\gamma_W=\gamma_n.
\end{equation}
This calculation can be fully duplicated to compute the spectral index of the centroid velocity map, in the case of a uniform density field. In this situation, 
the centroid velocity map (Eq.~\ref{eq_centroide2}) is just
\begin{equation}
C(x,y)=\beta \int v(x,y,z)\mathrm{d}z.
\end{equation}
We therefore have a similar result which reads
\begin{equation}
\gamma_C=\gamma_v.
\end{equation}

\clearpage


\clearpage


\clearpage

\clearpage

\clearpage

\clearpage

\clearpage

\clearpage

\clearpage

\clearpage

\clearpage

\clearpage

\clearpage

\clearpage

\clearpage

\begin{table}
\begin{center}
\begin{tabular}{c|cccc}
\tableline
\tableline
$\gamma_C$ & $\gamma_n=-3.0$ & $\gamma_n=-3.5$ & $\gamma_n=-4.0$ & $\gamma_n=-4.5$\\ \tableline
$\gamma_v=-3.0$ & -3.0 & -3.0 & -3.0 & -3.0\\
$\gamma_v=-3.5$ & -3.5 & -3.5 & -3.5 & -3.5\\
$\gamma_v=-4.0$ & -4.0 & -4.0 & -4.0 & -4.0\\ 
$\gamma_v=-4.5$ & -4.5 & -4.5 & -4.5 & -4.5\\ \tableline
\end{tabular}
\end{center}
\caption{\label{tab_centroid_exponent} Spectral index of the centroid velocity ($\gamma_C$) for the simulated PPV data cubes computed 
in the analysis. The uncertainty of the measured spectral indexes is 0.1.}
\end{table}

\begin{table}
\begin{center}
\begin{tabular}{c|cccc}
\tableline
\tableline
$\gamma_v$ & $\sigma(l$=1 pixel) & $l_{thin}(\Delta V=\delta v_{eff})$ \\ \tableline
-3.0 & 3 \kms & 0 pixel\\
-3.5 & 0.9 \kms & 4 pixels\\ 
-4.0 & 0.3 \kms & 24 pixels\\ 
-4.5 & 0.08 \kms & 42 pixels\\ \tableline
\end{tabular}
\end{center}
\caption{\label{tab_thinthick} These values were computed for $T=100$ K, $\delta u=0.25$ \kmsp, $\sigma(L)=3$ \kms and
L=129 pixels. The effective spectral resolution of our simulation (see Eq.~\ref{eq_delta_veff}) is 1.4 \kmsp.}
\end{table}

\end{document}